\documentclass[%
 reprint,
superscriptaddress,
longbibliography,
 aps]{revtex4-2}

\usepackage{dcolumn}
\usepackage{amsmath}
\usepackage{amssymb}
\usepackage{caption}
\usepackage{subcaption}
\usepackage{graphicx}
\usepackage{verbatim}
\usepackage{placeins}
\begin{document}

\preprint{}

\title{Chirality Amplification and Chiral Segregation in Liquid Crystals}
\author{Matthew J. Deutsch}
\affiliation{Advanced Materials and Liquid Crystal Institute, Kent State University, Kent Ohio, 44242, USA}
\author{Robin L. B. Selinger}
\affiliation{Advanced Materials and Liquid Crystal Institute, Kent State University, Kent Ohio, 44242, USA}
\affiliation{Physics Department, Kent State University, Kent, Ohio, 44242, USA}
\author{Paul van der Schoot}
\affiliation{Department of Physics and Science Education, Eindhoven University of Technology, 5612 AE Eindhoven, The Netherlands}

\date{\today}

\begin{abstract}
Liquid crystal mesophases of achiral molecules are normally achiral, yet in a few materials they spontaneously segregate and form right- and left-handed chiral domains. One mechanism that drives chiral segregation is molecular shape fluctuations between axial chiral conformations, where molecular interactions favor matching chirality and promote helical twist. Cooperative chiral ordering may also play a role in chirality amplification, as when a tiny fraction of chiral dopant drives a nematic phase to become cholesteric. We present a model of cooperative chiral ordering in liquid crystals using Maier-Saupe theory, and predict a phase diagram with a segregated cholesteric phase with alternating domains of left- and right-handed chiral twist, with opposite enantiomeric excess, in addition to racemic nematic and isotropic phases. Our model also demonstrates how chiral molecular fluctuations influence the helical twisting power of dopants in the nematic phase, which may be observed even in materials where the segregated cholesteric phase is preempted by a transition to another phase. We compare these results with Monte Carlo simulation studies of the switchable chiral Lebwohl-Lasher model, where each spin switches between right- and left-handed  chiral states. Simulation results validate the predicted phase diagram, demonstrate chiral amplification in the racemic nematic phase, and reveal complex coarsening dynamics in the segregated cholesteric phase. These results suggest that molecular fluctuations between degenerate chiral configurations may be a common mechanism to produce cooperative chiral order in achiral liquid crystals.
\end{abstract}

\maketitle

Chiral symmetry breaking and chirality amplification are attracting considerable interest in biology, chemistry, materials science and physics due to their relevance to the origin of life and single-handedness of molecules of biological origin~\cite{Blackmond2019}. In both, chirality at the molecular scale can be translated to the supramolecular and even the macroscopic scale~\cite{Tschierske2018}. For instance, a very small enantiomeric excess of chiral molecular building blocks produces helical polymers of definite handedness, not only in covalent polymers~\cite{Green1995} but also in supramolecular ones~\cite{vanGestel2006, Jouvelet2014}. Mixing tiny fractions of a chiral compound with achiral molecular building blocks does the same~\cite{Markvoort2011,Green1999}. 
Likewise, a small excess of one of the enantiomers or the addition of very small amounts of a chiral compound referred to as ``chiral doping'' is sufficient to break chiral symmetry in liquid crystals~\cite{Eelkema2006,Takezoe2012}.  Recently, lab-synthesized bacteria with biochemical components of opposite chirality to those observed in nature, so-called ``mirror cells", have been identified as a potential risk to life on Earth~\cite{Adamala2024}.

Generally, chirality amplification is thought to arise via {cooperative} ordering. In polymers, cooperative chiral ordering can occur via long-range correlations between left- and right-handed helically bonded states of neighboring molecular units along the backbone of a polymer chain~\cite{vanGestel2006, Selinger1996}. These long-range correlations arise because of an underlying Ising-like phase transition that in (quasi) one-dimensional systems is suppressed by thermal fluctuations.  
We argue here that via a similar mechanism, chiral interactions cause molecular conformations of nearby mesogens in a liquid crystalline fluid to become correlated or ``synchronized''~\cite{Tschierkske2016}. This cooperative chiral ordering of molecules gives rise to a diverging chiral susceptibility and must be at the root of chiral symmetry breaking in nematic liquid crystals. We show here that the helical twisting power associated with adding a chiral dopant to a nematic must be inversely proportional to the distance to a second-order phase transition, in which dynamic left-handed and right-handed conformers {phase separate} into cholesteric liquid crystals with opposite handedness. {Our calculations show that the helical twisting power may exhibit a complex temperature dependence.} 

Our work also explains how and why {a wide variety of} liquid crystalline materials consisting of achiral mesogens {can} exhibit {chiral segregation, also known as spontaneous resolution, to}  {form coexisting} right- and left-handed chiral domains~\cite{Pezl2002,Dierking2020}. 
{Although spontaneous resolution} has not been seen in ``conventional'' nematic liquid crystals {such as $n$-CB or MBBA, with the exception of, e.g., alkyl benzoic acid compounds~\cite{Strigazzi2001}}, {we put forward} that for such mesogens it may be obscured by a transition to a smectic or a crystalline phase. Even if obscured by a transition to another phase, {pre-transitional} effects should still be noticeable in the helical twisting power.
{Of particular interest for the present work is chiral segregation in fluids of achiral mesogenic compounds that exhibit a so-called nematic twist-bend phase~\cite{Tschierske2018,Majewska2022}. Like the cholesteric phase, this phase characterized by a helical modulation of the director field but with very much smaller pitch. Our theory provides an explanation for why such compounds are susceptible to chiral segregation.}

We first discuss the main ingredients and findings of our mean-field theory, and next discuss those of our Monte Carlo studies. We end with a discussion and conclusions, and suggestions for further study.

\subsection*{Methods}
\subsubsection*{Theory}
Following van der Meer \textit{et al}.~\cite{vandermeer1976}, we find from a mean-field analysis of the impact of a chiral Lebwohl-Lasher type of interaction potential between model mesogens to lead to expressions for the self-consistent fields experienced by species $\alpha=\pm$ due to the presence of species $\beta=\pm$ of the form
\begin{equation}\label{eq:selfconsistent_potential}
              V_{\alpha,\beta} (\theta) =  \left(-J(1-q^2) - q K_{\alpha,\beta}\right)S P_2(\cos \theta) f_{\beta}. 
\end{equation}
The chiral coupling constants $K_{\alpha,\beta}$ are $K_{+,+} = K$ between the $+$ species, $K_{-,-} = - K$ between the $-$ species and $K_{+,-}=K_{-,+}=0$ between the species with opposite handedness. In Eq.~\ref{eq:selfconsistent_potential}, we have absorbed unimportant numerical constants in the definitions of the interaction strengths $J$ and $K$ as well as the cholesteric wave number $q$ that in our description is made dimensionless by multiplication with a microscopic length scale~\cite{vandermeer1976}. The free energy of interaction follows from summing all self-consistent fields experienced by species $\alpha$ multiplied with the appropriate fraction $f_\alpha$, and accounting for double counting. Adding the Gibbs entropy for the orientational distribution and that for the distribution over the species $\alpha$, as well as the contribution of the biasing field $g$, then produces, after some rearranging, Eq.~\ref{eq:free_energy}.

\subsubsection*{Simulation Methods}
To explore cooperative chiral ordering and spontaneous chiral segregation by means of Monte Carlo computer simulation, we study a switchable chiral 3D Lebwohl-Lasher model based on the Hamiltonian of Eq.~\ref{eq:MC_hamiltonian_gfield}. 
We perform Monte Carlo simulations of the model via the Metropolis algorithm, where at each Monte Carlo step we separately vary the local orientation $\hat{u}_i$ and chiral state parameter $\eta_i$. We use a novel method for trial spin reorientation which realizes a smoother reorientation of the spin, rather than the Barker-Watts technique~\cite{barker1969} used in~\cite{skacej2021}. To perform a random walk on a unit sphere, we calculate two vectors orthogonal to the orientation vector $\hat{u}_i$ and tangent to the unit sphere, then reorient the spin to a random position on the edge of a disk of radius $d$, and normalize the orientation vector to unit length. The disk size $d$ represents the step size and is adjusted to keep the Monte Carlo acceptance rate in the range of 0.4 to 0.6 during equilibration at each temperature, and after that is held fixed. We initialize each simulation in the racemic isotropic phase.

Instead of periodic boundary conditions, we use open boundary conditions. This allows twist without any constraints imposed by periodic boundary conditions, and does not impose any preferred orientation of the twist axis. We find that the spontaneously chosen twist axis is typically \emph{not} along the $x, y$ or $z$ axis, as in fact was assumed in~\cite{elsasser2022}, but along a body diagonal of the cubic lattice. We also note that open boundary conditions allow topological defects to nucleate and annihilate at the free surfaces.

Lattice-based Monte-Carlo simulations are a computationally ``embarrassingly parallelizable'' problem.
Our Monte Carlo code is highly optimized to run on graphics processing units (GPUs) using a combination of Fortran and OpenACC for the bulk of the computation, and Python for simulation management, data processing, analysis, and visualization. Simulations were performed on the National Center for Supercomputing Applications' Delta cluster through the NSF ACCESS program~\cite{access}. A copy of the simulation code is available~\cite{zenodo2025}.

\subsection*{Results}
\subsubsection*{Maier-Saupe Theory}
Our starting point is the Maier-Saupe theory of van der Meer and collaborators~\cite{vandermeer1976} that we extend to mixtures of $\pm$ enantiomers: compounds with axial chiralities P and M, and  opposite cholesteric pitches. The fraction of $\pm$ components $f_+=1-f_{-}$ we treat as an equilibrium variable. For example, in non-chiral biphenyl mesogens {such as, say, 5CB} these stereo isomers may inter convert dynamically~\cite{Eelkema2006} with estimated barriers a few times the thermal energy~\cite{Zannoni2022}. Interactions between equal pairs of $\pm$ components favor a dimensionless helical wave number $\pm q$. In this model, interactions between $+$ and $-$ components do not produce a helical pitch{, and hetero-chiral interactions are less strong than homo-chiral ones}. For simplicity all nematic interactions between the various components are equal, and described by a single coupling constant $J>0$. In contrast to the earlier work~\cite{Shiyanovskii1992}, our model explicitly deals with chiral interactions between the mesogens and predicts a cholesteric pitch. Our Monte Carlo simulations, discussed in more detail in the next section, are based on the same type of interaction potential~\cite{vandermeer1976}. 

Accounting for the mixing entropy of the two species of compound, we obtain for the Helmholtz free energy $F$ per mesogen
\begin{equation}\label{eq:free_energy}
\begin{split}
    F & = k_\mathrm{B} T \langle \ln P \rangle - \frac{1}{2}JS^2  \\ & +\frac{1}{2}\left(Jq^2-Kq\right)S^2f_+^2  + \frac{1}{2}\left(Jq^2+Kq\right)S^2f_-^2 \\ & + k_\mathrm{B} T f_+\ln f_+ + k_\mathrm{B} T f_-\ln f_-  - 2 g f_+,
\end{split}
\end{equation}
where $k_\mathrm{B} T$ denotes the thermal energy with $k_\mathrm{B}$ Boltzmann's constant and $T$ the absolute temperature. The first two terms represent the classical expression for the Maier-Saupe free energy (per molecule) describing the isotropic-nematic phase transition~\cite{vanderschoot2022}. The first describes the Gibbs entropy associated with the distribution $P=P(\cos \theta)$ over the polar angle $\theta$ between a mesogen and the local director, and the second the contribution from the nematic interaction between the mesogens. In the latter, $S = \langle P_2(\cos \theta) \rangle \in[-1/2,+1]$ denotes the scalar nematic order parameter taken to be the same for both $\pm$ components, with $P_2(\cos \theta)=\frac{3}{2}(\cos \theta)^2 -\frac{1}{2}$ the second Legendre polynomial. The angular brackets $\langle \cdots \rangle$ indicate an orientational average $\int_{-1}^{+1} d\cos \theta \hspace{0.1cm} P(\cos \theta)(\cdots)$ over the orientational distribution function $P(\cos \theta)$ of the nematogens. We ignore  any biaxility in the distribution function that may arise in the cholesteric phase.

The third and fourth terms in Eq.~\ref{eq:free_energy} arise due to the chiral interaction that induces a helical deformation of the director field with dimensionless wave number $q$. The quadratic terms in $q$ account for the reduction of the nematic interaction in the twisted director field, while the terms linear in $q$ measure how strongly the chiral interaction is enhanced.  
These terms follow from a mean-field analysis of a chiral Lebwohl-Lasher type of interaction potential between model mesogens~\cite{vandermeer1976}; see also Eq.~\ref{eq:MC_hamiltonian_gfield}. 

The last three terms describe the mixing entropy of the species $\pm$, and a term that breaks chiral symmetry with $g$ a free energy difference between the two enantiomers. This chiral biasing term may, e.g., represent a chemical potential difference that in a grand canonical ensemble regulates the enantiomeric excess $\eta = 2f_+-1 = 1-2f_-$. Alternatively, it may be a free energy difference between the two enantiomers induced by the action of circularly polarized light or the interaction with a chiral dopant~\cite{Eelkema2006}. In the latter case we expect that $g$ is proportional to the concentration of dopant, the consequences of which we investigate in more detail below. {See the SI for a discussion.}

The optimal orientational distribution function $P(\cos \theta )$ minimizes the free energy, as do the equilibrium values of the chiral wave vector $q$ and the fraction $+$ and $-$ states $f_+=1-f_-$. For $g=0$ and a sufficiently weak chiral interaction, the transition between a racemic isotropic and a racemic nematic phase occurs at the clearing temperature $T_\mathrm{IN}$ for which $J/k_\mathrm{B}T_\mathrm{IN} = 4.54$ where $S=0.43$ in the nematic phase~\cite{vanderschoot2022}, when the free energies of the coexisting isotropic and nematic phase are equal. Let us consider the fluid to be in the nematic phase, with temperature $T< T_\mathrm{IN}$ and order parameter $S>0.43$. 

If we minimize the free energy Eq.~(\ref{eq:free_energy}) with respect to $q$, we obtain 
\begin{equation}\label{eq:optimal_wave_number} 
    x\equiv \frac{q}{q_0}=\left(\frac{f_+^2-f_-^2}{f_+^2+f_-^2}\right),
\end{equation}
with $q_0 = K/2J$ the intrinsic cholesteric wave number of the enantiomers and $x$ the ratio of the equilibrium value of $q$ and $q_0$. We require that $|q_0| \lesssim 1$ or $|K|/J \lesssim 2$ for the pitch not to drop below a microscopic length.

We also need to minimize the free energy Eq.~\ref{eq:free_energy} with respect to $f_+ = 1-f_-$, and find
\begin{equation}\label{eq:optimal_fraction}
\begin{split}
    \chi  \left(\frac{1}{2}x^2-x\right) f_+ - \chi \left(\frac{1}{2}x^2+x\right) f_- \\ +\ln \left(\frac{f_+}{f_-}\right) -  \frac{2g}{k_\mathrm{B} T} = 0,
    \end{split}
\end{equation}
where the strength of the chiral interaction is given by $\chi \equiv (K/J)^2S^2J/2k_\mathrm{B}T = 9.08 \hspace{0.1cm} q_0^2 \hspace{0.1cm} S^2 \hspace{0.1cm} (T_\mathrm{IN}/T) $. The temperature dependence of $S=S(T/T_\mathrm{IN})$ follows from the self-consistent field equation of Maier-Saupe theory~\cite{vanderschoot2022}. In the absence of a chiral interaction between the mesogens, $\chi =0$. This happens if $K = 0$ and $q_0=0$, or if the fluid is in the isotropic phase and $S = 0$. The fractions $\pm$ enantiomers then obey a simple Boltzmann distribution, giving for the enantiomeric excess  $\eta=\tanh g/k_\mathrm{B} T$, not unlike the average spin state of the Ising model of ferromagnetism where $g$ takes on the role of magnetic field. In our model, the ``spins'' do not directly couple, only indirectly -- see Eq.~\ref{eq:MC_hamiltonian_gfield}.

\begin{figure}
    \centering
    \includegraphics[width=\linewidth]{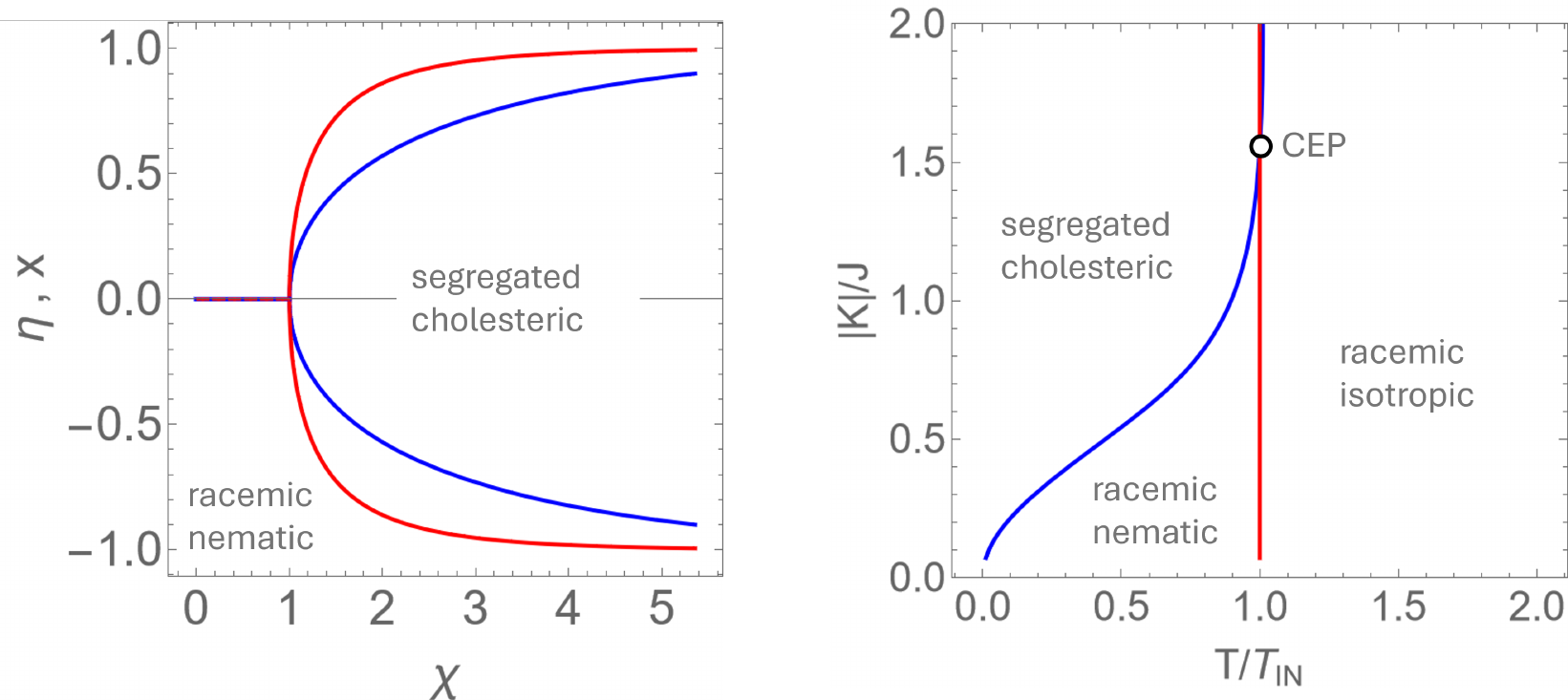}
    \caption{Left: Enantiomeric excess $\eta$ (blue -- inner curve) and ratio $x$ of the cholesteric wave number $q$ and the maximal value $q_0$ (red -- outer curve) in the co-existing {segregated} cholesteric phases as a function of the dimensionless magnitude of the chiral interaction between the nematogens $\chi$. Right: phase diagram as a function of the temperature $T/T_\mathrm{IN}$ scaled to the nematic transition temperature $T_\mathrm{IN}$, and the ratio of the chiral and nematic interaction strengths $|K|/J$. The vertical line in red indicates the isotropic-nematic transition. The curved line in blue demarcates the chiral transition between the {segregated} cholesteric phase and the racemic nematic phase for $T<T_\mathrm{IN}$ and the racemic isotropic phase for $T>T_\mathrm{IN}$, {which cross at what we interpret to be a critical end point (CEP) indicated by the circle.} See also the main text.} 
    \label{fig:phasediagram}
\end{figure}

For $g=0$, there cannot be an overall enantiomeric excess as the mesogens have no preference for either enantiomeric state. However, \emph{local} chiral symmetry may still be broken. This happens if $\chi >1$, as is shown in Fig.~\ref{fig:phasediagram} (left), having numerically solved Eqs.~\ref{eq:optimal_wave_number} and \ref{eq:optimal_fraction} for $g=0$. 
For $\chi \leq 1$ the enantiomeric excess $\eta \equiv 2f_+-1$ and cholesteric wave number $x = q/q_0$ are both equal to zero. This represents the achiral, racemic nematic phase. For $\chi \geq 1$, local chiral symmetry is broken and the racemic nematic phase {segregates} into two co-existing cholesteric phases with opposite cholesteric wave number $\pm |x|$ and opposite enantiomeric excess $\pm |\eta|$. In our model, the chiral or {mirror-image symmetry breaking} transition is continuous and is reminiscent of the spontaneous magnetization {of a ferromagnet} below the Curie temperature.

The condition {$\chi_\mathrm{c} = 9.08 \hspace{0.1cm}q_0^2 \hspace{0.1cm}S^2_\mathrm{c} \hspace{0.1cm} T_\mathrm{IN}/T_\mathrm{c} = 1$} identifies a critical chiral transition or {chiral segregation} temperature $T_\mathrm{c}/T_\mathrm{IN} = 9.08 \hspace{0.1cm} q_0^2 S^2_\mathrm{c}$ with $S_\mathrm{c}$ the nematic order parameter at $T=T_\mathrm{c}$. For weak chiral interaction, implying small values of $q_0=|K|/2J$, $T_\mathrm{c} \ll T_\mathrm{IN}$. {In that case} the chiral transition might be obscured by a transition to a smectic or a crystalline phase. On the other hand, $T_\mathrm{c}$ moves towards $T_\mathrm{IN}$ with increasing chiral interaction strength, as is shown in Fig.~\ref{fig:phasediagram} (right). The chiral transition crosses the nematic transition at for $|K|/J = 1.54$. For $|K|/J > 1.54$ there is a direct transition from a racemic isotropic phase to two {segregated} cholesteric phases of {opposite handedness and enantiomeric excess, and} equal volume. {This arguably marks a critical end point. (See also below.)} The phase diagram we obtain from our simulations, Fig.~\ref{fig:sim_phase_diag} (top), is in good agreement with that from our mean-field theory, Fig.~\ref{fig:phasediagram} (right).

Not surprisingly, we obtain mean-field critical exponents near the chiral transition. For $T \rightarrow T_\mathrm{c}$, Eqs.~\ref{eq:optimal_wave_number} and \ref{eq:optimal_fraction} can be solved exactly, giving $x\sim \pm 4 \hspace{0.1cm } (3/28)^{1/2}(\chi-1)^{1/2} \pm \propto(T_\mathrm{c}-T)^{1/2}$ and $\eta \sim \pm 2 \hspace{0.1cm} (3/28)^{1/2}(\chi-1)^{1/2}\propto \pm(T_\mathrm{c}-T)^{1/2}$ for $\chi \geq 1$ or $T\leq  T_\mathrm{c}$, and $\eta = x = 0$ for $\chi \leq 1$ or $T> T_\mathrm{c}$. 
Arguably, the existence of a second order chiral transition explains the phenomenon of chirality amplification because {thermodynamic} susceptibilities tend to become large and even diverge {upon approach of} the critical point. This shows up in the response of the chiral order parameters $x$ and $\eta$ to the chiral biasing field $g$. 

\begin{figure}
    \centering
    \includegraphics[width=\linewidth]{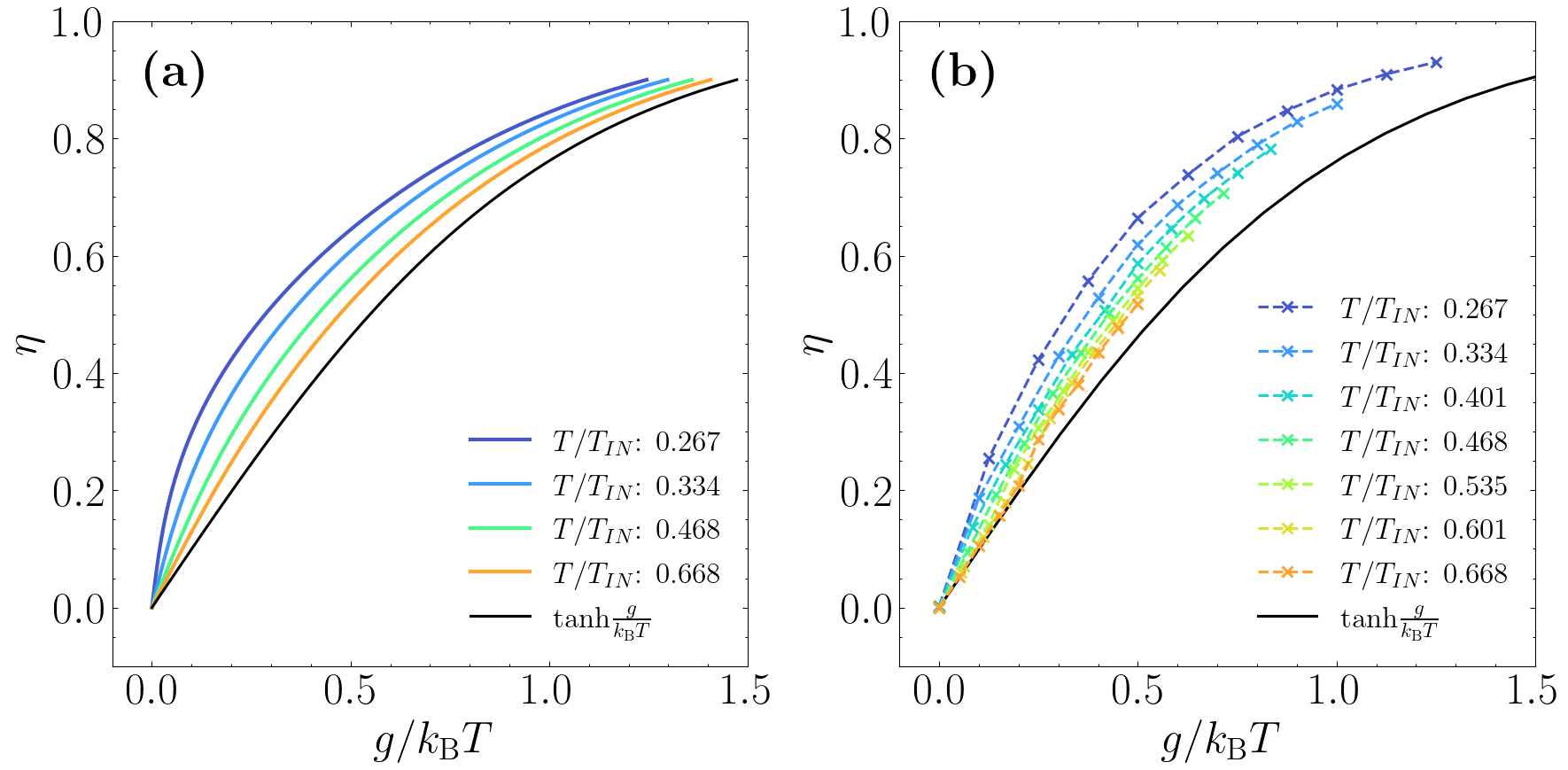}
    \caption{Enantiomeric excess $\eta$ for a fixed strength of the chiral interaction $K/J = 0.333$, as a function of the dimensionless chiral biasing potential $g/k_\mathrm{B}T$ for different temperatures $T/T_\mathrm{IN}$. (a) Predictions from Maier-Saupe theory. The corresponding values of the chiral interaction strength $\chi$ are $0.825, 0.634, 0.412$ and $0.239$ from top to bottom. (b) Results of Monte Carlo simulations. Indicated is also the ideal hyperbolic tangent behavior expected in the absence of a chiral interaction and $K=0$.}
    \label{fig:chirality_amplification}
\end{figure}

This point is illustrated in Fig.~\ref{fig:chirality_amplification} (a). The figure shows our solution to Eqs.~\ref{eq:optimal_wave_number} and \ref{eq:optimal_fraction} for the enantiomeric excess $\eta$ as a function of $g/k_\mathrm{B}T$ for different scaled temperatures $T/T_\mathrm{IN}$ and fixed chiral interaction $K/J=0.333$. Clearly, the enantiomeric excess $\eta$ turns out not only a function of the dimensionless biasing field $g/k_\mathrm{B} T$ but also of $T/T_\mathrm{IN}$. The reason is that the chiral interaction strength $\chi$ is a function not only of $K/J$ but also of $T/T_\mathrm{IN}$.

{For the temperatures $T/T_\mathrm{IN}$ shown, $\chi$ is smaller than unity and the fluid remains in the racemic nematic phase.} Still, we do see a significant increase in the enantiomeric excess relative to the hyperbolic tangent relation describing the case $K=0$, and more so the lower the temperature. 
The trends of our theoretical predictions agree with those from our simulations, as Fig.~\ref{fig:chirality_amplification} (b) confirms. See also the Supporting Information, Fig.~S1, in which we investigate the impact of larger and smaller values of $K/J$. For $\chi>1$, the Maier-Saupe theory predicts hysteresis of the enantiomeric excess $\eta$ as a function of the biasing field $g$, with metastable positive (negative) valued $\eta$ for negative (positive) values of $g$ {(shown in SI)}. 
 
In the limit $\chi \rightarrow 1$ the susceptibilities of the enantiomeric excess $\chi_\eta \equiv \lim_{g\rightarrow 0} \partial \eta / \partial g $ and of the cholesteric wave number $\chi_x \equiv \lim_{g\rightarrow 0} \partial x / \partial g$ both diverge. From Eqs.~\ref{eq:optimal_wave_number} and \ref{eq:optimal_fraction}, we find that for $\chi \uparrow 1$, $\chi_\eta = \frac{1}{2} \chi_x = \frac{1}{4} (1-\chi)^{-1}/k_\mathrm{B}T$, while for $\chi \downarrow 1$, $\chi_\eta = \frac{1}{2} \chi_x  \sim \frac{7}{24} (\chi -1)^{-1}/k_\mathrm{B}T$. So, in both cases we find $\chi_\eta \propto |T-T_\mathrm{c}|^{-1}$. Our simulations produce {approximately} the same mean-field exponent (Fig.~S4) although we have not performed a finite-size analysis to obtain a more accurate value. Finally, accounting in the free energy for square gradient terms in the variables $x$ and $f_+ = (\eta+1)/2$ produces two correlation lengths proportional to $1/\sqrt{1-\chi}$ confirming that chiral conformational states are indeed in a way contagious and {transferred to nearby} mesogens through the chiral interaction.

The susceptibility $\chi_x$ associated with the cholesteric wave number may be interpreted to represent what is known as the helical twisting power of a chiral dopant~\cite{Ferrarini1996}. 
A tiny induced preference for a particular enantiomeric form of the nematogens becomes strongly enhanced by the coupling to the chiral symmetry breaking transition. This may well explain why so little chiral dopant is needed to turn a nematic into a cholesteric~\cite{Eelkema2006,Takezoe2012}. 
{Differences in the helical twisting power associated with different dopants} may be understood in terms of unequal surface interactions with the director field~\cite{Ferrarini2009}, expressed in our model in different values of the interaction free energy, $g$. In fact, {as we show in the SI} for large enough binding {free} energies, $g$ may be replaced by $g X$ with $X\ll 1$ the mole fraction of dopant, in which case the helical twisting power, $\beta$, can be written as 
\begin{equation}\label{eq:helicaltwistingpower}
\beta = \frac{1}{1-\chi} \left(\frac{2g}{k_BT}\right) \frac{q_0}{2\pi}.
\end{equation}

The helical twisting power efficacy of the same dopant in different nematogens may be interpreted in terms of the distance in temperature to the chiral transition of that nematogen, yielding different values of $\chi$, and differences in the binding free energy, $g$, as well as the intrinsic helical wave number, $q_0$. Changes in the binding free energy between different dopants result in different helical twisting powers dependent on the nematogen. Novel in our expression for the helical twisting power Eq.~\ref{eq:helicaltwistingpower} is the pretransitional enhancement factor $(1-\chi)^{-1}$ not accounted for in other approaches~\cite{Ferrarini2009}. {See also the SI.} 

The temperature dependence of the helical twisting power {that we predict is \textit{not} universal}, as $\chi$, $g$ and $q_0$ all depend on the temperature. For our model, $q_0=K/2J$ is often thought to not depend on temperature~\cite{vandermeer1976} but this presumes that $K$ and $J$ are energies. {However}, because the interaction energy {is that of} a coarse-grained model {in which degrees of freedom have been ``integrated out''}, $K$ and $J$ must in fact be interpreted as free energies {that are} temperature dependent. If we ignore this and presume $K$ and $J$ to be a constant of temperature, we deduce from Eq.~\ref{eq:helicaltwistingpower} that $\beta$ should decrease with temperature under most conditions because higher temperatures imply larger distance to the critical {chiral segregation} temperature and smaller values of $\chi$. This is indeed usually seen experimentally, but not always~\cite{Kitzerow2001}. However, if the binding of the dopant to the nematogen is entropy- rather than enthalpy-{dominated}, and $g/k_BT$ increases with temperature, then $\beta$ may actually \emph{increase} with temperature. This could explain recent observations~\cite{Kikuchi2019}. We refer to the SI for a more detailed discussion of the factors that influence the temperature dependence of the helical twisting power $\beta$. 

\subsubsection*{Simulation Results}
To explore cooperative chiral ordering and spontaneous {chiral segregation}, we study by means of Monte Carlo simulation a switchable chiral Lebwohl-Lasher model as described by the following Hamiltonian:
\begin{multline}
    H = \sum_{\langle i,j \rangle} \left[ - J \hspace{0.05cm} P_2\left(\hat{u}_i \cdot \hat{u}_j\right) \right. \\
    \left. -\frac{1}{2} K \left [ (\hat{u}_i \times \hat{u}_j ) \cdot \hat{r}_{ij} \right] (\hat{u}_i \cdot \hat{u}_j )(\eta_i + \eta_j) - g \eta_i \right],
    \label{eq:MC_hamiltonian_gfield}
\end{multline}
where the sum is over neighboring pairs $\langle i,j \rangle$ of lattice sites, $J$ and $K$ have the same meaning as before, as does the biasing field $g$, and $\hat{r}_{ij}$ denotes the unit vector connecting these two sites. Associated with each site are two order parameters: the unit orientation vector $\hat{u}_i$ and the chirality $\eta_i=\pm 1$ describing the left or right handedness of the mesogens. The first term in Eq.~\ref{eq:MC_hamiltonian_gfield} represents the usual Lebwohl-Lasher potential~\cite{lebwohl1972,skacej2021}, and the second term is a chiral interaction term {which was} recently explored in a different context by Elsasser and Kuhnhold~\cite{elsasser2022} without the local switchable chirality part of our model. {Our model incorporates a} coupling term $\eta_i + \eta_j${, promoting}
twist between neighboring ``spins'' for pairs of like chirality $\eta_i =\eta_j$ but not between neighboring spins of opposite chirality $\eta_i\neq \eta_j$. More simulation details are available in the {Methods} section. 
Simulations were performed on the National Center for Supercomputing Applications' Delta cluster through the NSF ACCESS program~\cite{access}. A copy of the simulation code is available at~\cite{zenodo2025}.

\begin{figure}
    \centering
    \includegraphics[width=0.849\linewidth]{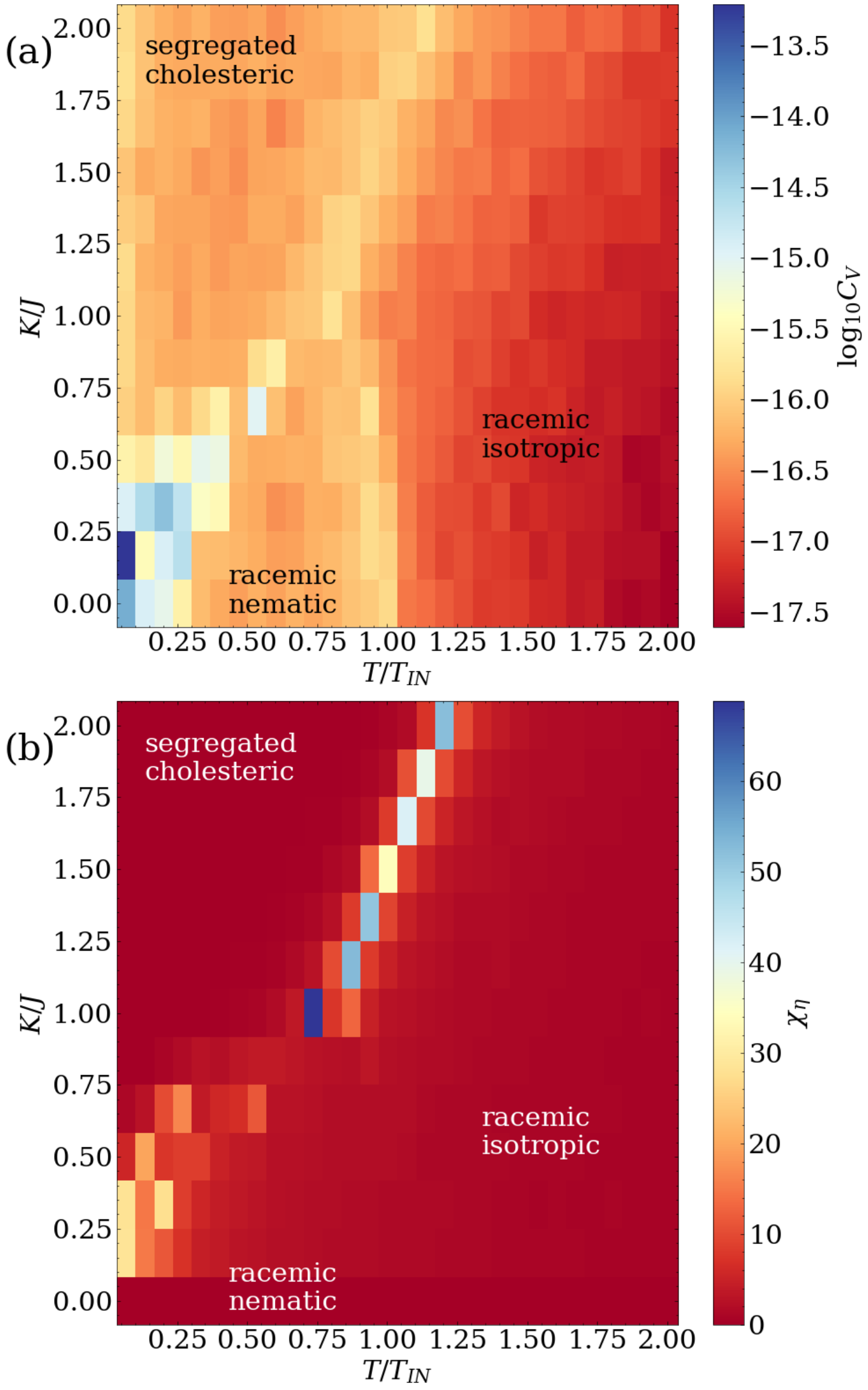}
    \caption{Phase diagrams as a function of the chiral interaction strength $K/J$ and scaled temperature $T/T_\mathrm{IN}$ obtained by means of Monte Carlo simulations. (a) Heat map of the logarithm of the specific heat, $\mathrm{log}_\mathrm{10} C_V$, and (b) that of the chiral susceptibility $\chi_\eta$. The latter cannot distinguish between the racemic isotropic and nematic phases. 
       }
        \label{fig:sim_phase_diag}
\end{figure}

We simulate the model {on} an $N \times N \times N$ cubic lattice with $N=128$. To explore the phase diagram, the chiral interaction strength, $K/J$, was varied from $K/J = 0.0$ to $K/J = 2.0$ with a total of $13$ values.  
For each value of $K/J$, the temperature was varied from $T/T_{IN} = 2.0 $ to $T/T_{IN} = 0.067$ with $2 \times 10^6$ Monte Carlo steps at each of the $30$ temperature values. For each set of parameters, we used the last $6.25\% $ of configurations to calculate specific heat and chiral susceptibility, with configurations separated by 100 Monte Carlo steps per spin. The specific heat was calculated from the energy fluctuations as $C_V = \left( \langle E^2 \rangle - \langle E \rangle ^2 \right) /k_\mathrm{B}T^2$. The
chiral susceptibility is calculated from the fluctuations in the enantiomeric excess $\eta$, $\chi_\eta = N^3 \left ( \langle \eta^2 \rangle - \langle \eta \rangle ^2 \right)/k_\mathrm{B} T $. The enantiomeric excess was calculated as the average chirality per lattice site, in direct analogy to the average spin state in the Ising model.

The phase diagram obtained from our simulations is shown in Fig.~\ref{fig:sim_phase_diag} (a), showing peaks in the heat capacity that indicate transitions between racemic isotropic, racemic nematic and {segregated} cholesteric phases {with opposite handedness and enantiomeric excess}, in close agreement with predictions of our Maier-Saupe theory of Fig.~\ref{fig:phasediagram}. We find a racemic isotropic phase at high temperature, a racemic nematic phase at low temperature and sufficiently weak chiral interaction $K/J$, and spontaneously {segregated} cholesteric domains at low temperature and high $K/J$. {The intersection of the phase boundaries points at the existence of a critical end point}, also predicted by our Maier-Saupe theory. {In addition, we} plot the chiral susceptibility in Fig.~\ref{fig:sim_phase_diag} (b), which can only show the phase boundary where {spontaneous resolution takes place}. It confirms that near the transition, chiral fluctuations become large and hence also the chiral susceptibility {that translates to a large helical twisting power. Small differences in the critical exponent for the chiral susceptibility that we find for temperatures near the critical end point and much below that indicate that we cannot rule out that it may be tricritical point. See the SI.}.

Even not all that close to the transition, the increased susceptibility can be quite significant as Fig.~\ref{fig:chirality_amplification} (b) confirms. In the figure, the enantiomeric excess $\eta$ is shown as a function of the dimensionless chiral bias field for different temperatures but fixed chiral interaction strength $K/J=0.333$. In our Monte Carlo simulations, we varied $g/k_\mathrm{B}T$ from $0.0$ to $0.25$, $K/J$ from $0$ to $0.5$, and $T/T_{IN}$ from $ 0.668$ to $0.267$, where we annealed $2 \times 10^6$ Monte Carlo steps at each temperature. In agreement with the theory, we find that the chiral susceptibility goes up with decreasing temperature and with increasing chiral interaction strength $K/J$, as the Fig.~S1 shows. According to our theory, both $T$ and $K/J$ enter because $\eta$ not only depends on $g/k_\mathrm{B}T$ but also on the value of $\chi$ that itself is a function of $T/T_\mathrm{IN}$ and $K/J$.

\begin{figure}
    \centering
    \includegraphics[width=\linewidth]{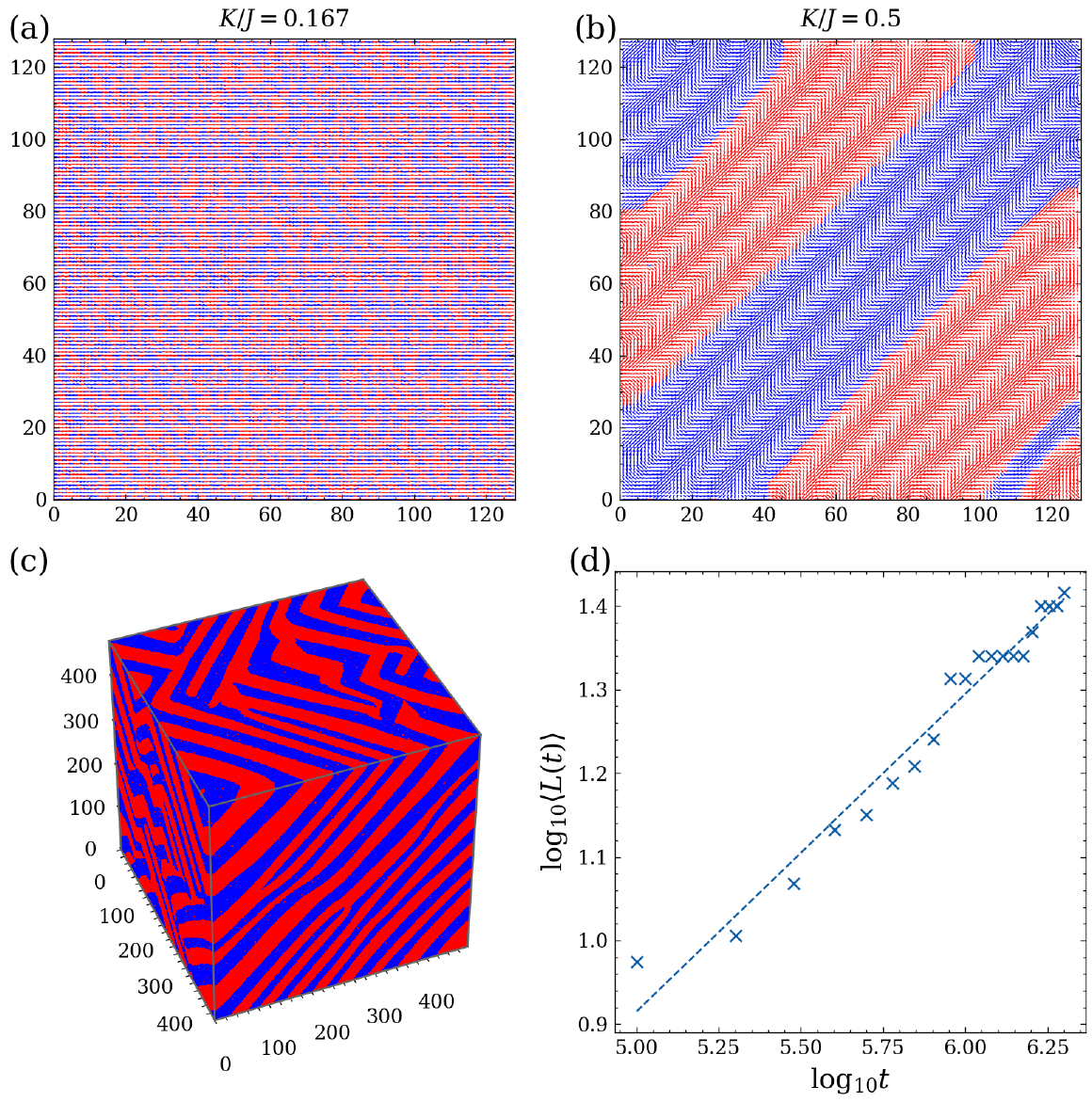}
    \caption{(a-b): Simulation results showing local orientation and chirality configurations at fixed temperature $T/T_{IN} =0.067$ and varied values of chiral interaction, $K/J$. A horizontal $XZ$ cross-section is shown with the local unit orientation field $\hat{u}_i$ colored blue and red corresponding to $\eta_i=\pm1$. Simulation box size $N = 128$.  
    (c): Simulation of size $N = 480$ after $2\times 10^6$ steps showing alternating chiral domains. (d): Double-logarithmic plot of the average layer thickness, $\langle L(t) \rangle$, as a function of time, $t$, in number of steps for the box size $N = 480$. Stripes of alternating chirality where counted across the $101$ plane every $1\times 10^5$ steps. The dashed line shows a curve fit of $\langle L \rangle \propto t^\alpha $ producing a scaling exponent $\alpha = 0.379 \pm 0.016$.
        }
    \label{fig:4}
\end{figure}

What Maier-Saupe theory cannot show, and our simulations can, is the spatial structure of the phases and their coarsening behavior. Fig.~\ref{fig:4} (a) and (b) show $xz$ cross-sections of a system with size $N=128$, with open boundary conditions on all sides. Both simulations were annealed to a temperature of $T/T_{IN} =0.067$ starting from $T/T_{IN} = 1.0$, in $30$ steps of $2 \times 10^6$ Monte Carlo steps each. In the figures, right-handed ${\eta_i}=+1$ spins are shown in red and left-handed ${\eta_i}=-1$ spins are shown in blue. Fig.~\ref{fig:4}(a) shows the racemic nematic phase at low value of $K/J=0.167$, with short range chiral correlations, but no long range order and no induced cholesteric twist.
Fig.~\ref{fig:4}(b) is at a higher value of $K/J=0.5$, showing {coexisting} layers of {left- and right-handed} cholesterics  approximately two pitch lengths each. A more detailed image can be found in Fig.~S3,  demonstrating that under these conditions the interface between left- and right-handed cholesteric layers is very sharp. The {cholesteric} domains form with the pitch axis oriented along a body diagonal of the cubic lattice, see also Fig.~\ref{fig:4} (c). This orientation of the pitch axis is allowed by open boundary conditions and is lower in energy than a pitch oriented along the $x$, $y$ or $z$ axes of the cubic simulation volume.

To investigate coarsening behavior in the {mirror-image symmetry broken} phase, we carried out a simulation of a larger system with $N=480$ and open boundary conditions, for a fixed chiral interaction ${K/J}=0.5$ and a temperature $T/T_{IN} = 0.067$ that falls in the {segregated} cholesteric region of the phase diagram. We followed the coarsening over $2 \times 10^6$ Monte Carlo steps per spin. Our findings are shown in Fig.~\ref{fig:4} (c) and  (d), where layers with ${\eta_i}=1$ are again shown in red and with ${\eta_i}=-1$ in blue. Layers with alternating chiral sense form within the first $1 \times 10^5$ {Monte Carlo} steps after starting from an initial random configuration as seen in Fig.~S5 (a). {The sharp interface between left- and right-handed cholesteric layers is apparent even in these early configurations as shown by Fig.~S7.} These layers widen with time, shown in more detail in the cross sections shown in Fig.~S2 and Fig.~S5. Initially, multiple domains form with layer normals aligned along different body diagonal axes. Over time, those domains grow, the average layer width increases, and the number of layers drops. Layers annihilate via two mechanisms: (i) motion and annihilation of edge dislocations in the layer structure, and (ii) layer annihilation at corners of the cubic box facilitated by the open boundary conditions.
The average layer thickness $L$ we find to increase continuously as $L \sim {t^{\alpha}}$ with scaling exponent $\alpha= {0.379} \pm {0.016}$ as shown in Fig.~\ref{fig:4} (d). Because layer formation and coarsening are driven by an intrinsically anisotropic surface tension, we do not expect this scaling exponent to be universally valid~\cite{Vollmayr1999,Sethna2023}. 

As Fig.~\ref{fig:4} (c) shows, the domain structure is quite complex, each with layer normals oriented along different body diagonals. Layers formed in early stages of pattern formation have thicknesses less than the cholesteric pitch, increasing to several pitch lengths during coarsening. See Fig.~S3, Fig.~S6 b), and Fig.~S7. Layers less wide than about a pitch do not have a simple, single twisted cholesteric structure but exhibit a double-twist structure. The system remains in a poly-domain structure after $2 \times {10^6}$ Monte Carlo steps per spin. The dynamics of micro-structural evolution is shown in Supplementary Video S1 and in Fig.~S5. Cross-sectional images {over the entire simulation length} are shown in Fig.~S2 {with selected insets shown in Fig.~S7}.

Finally, for values of $K/J=0.833$ and larger, a more complex, disordered double twist geometry without a uniform twist axis emerges{, as seen in Fig.~S6 (d)}. The defect-rich micro-structures suggest the formation of blue phases, which have not previously been observed in studies of the chiral Lebwohl-Lasher model. We plan to investigate the details of the blue phase in the uniform chirality Lebwohl-Lasher model in future work.

\subsection*{Discussion and Conclusion}
{Chiral segregation and deracemization (conversion of a racemate to a mixture with enantiomeric excess) were first observed in racemic mixtures of chiral molecules during crystallization, but can also arise in a variety of other material systems~\cite{Buhse2021}.
Indeed, a wide variety of achiral mesogens are known to segregate in chiral domains in different liquid-crystalline phases~\cite{Pezl2002,Tschierkske2016}. For instance, bent-core molecules self-assemble into smectic phases with a chiral arrangement of layer normals, molecular dipole orientations, and molecular tilts, and chemically linked ``multimeric'' mesogens as well as ferronematic compounds spontaneously segregate in oppositely handed helicoidal twist-bend nematic domains~\cite{Link1997,Gortz2005,Prasang2008,Majewska2022,Karcz2024}.}

Current theoretical understanding of why achiral mesogens break mirror-image symmetry is based, on the one hand, on macroscopic elasticity theory~\cite{Dozov2001,Jakli2018}, and, on the other, on results of computer simulations that focus in particular on the role of particle shape~\cite{Memmer2002,Wilson2018,Chiappini2019}. The role that configurational fluctuations might have that produce degenerate right- and left-handed twisted conformations has attracted much less attention in the theoretical literature~\cite{Takezoe2012}. Such chiral fluctuations might become spatially correlated~\cite{Tschierkske2016}, and drive chiral segregation~\cite{Shiyanovskii1992,Debenedetti2023}. Wilson \textit{et al.}~have clearly demonstrated the ``contagiousness'' of chiral conformational states via atomistic molecular simulations of flexible bent-core mesogens~\cite{Earl2005,Wilson2018}. Our combined mean-field theory and computer simulation study demonstrates that this mechanism can play a key role in driving chiral segregation and giving rise to the large helical twisting power typical of chiral dopants.

The mean-field theory and computer simulations that we have presented confirm that favorable homo-chiral interactions between nematogens cause their chiral conformational states to become correlated and promote segregation in oppositely twisted cholesteric domains. That the predicted mirror-image symmetry transition is a continuous one provides an explanation for the large helical twisting power that chiral dopants can have, even if the transition is hidden, e.g., by a transition to a crystal state (see the SI). 

Our Maier-Saupe theory predicts $T_\mathrm{c}/T_\mathrm{IN} = 9.08 \hspace{0.1cm} q_0^2 \hspace{0.1cm} S^2_\mathrm{c}$ for the chiral segregation temperature. Hence, to observe the chiral segregation in the nematic phase, we need this ratio to be close enough to unity. This implies that the pitch of the pure P enantiomer must be $P_0 = 2\pi/q_0 \approx 10-20 $ in units of microscopic length. If we take for this the particle length, then this is an unlikely small number with cholesterics in mind. Helicoidal twist-bend nematics, however, are known to have pitches of this order of magnitude, presumably due to their much larger core and hence stronger chiral interaction. Arguably, this explains why chiral segregation is more often seen in nematogens that support this phase.

Finally, our model presumes the P and M configurations of the nematogen to be the most likely configurations, which for $n$-CB certainly is the case. This need not be for other nematogens, which may have as equally stable or more stable configurations achiral ones. For instance, the nematogen DIO exhibits chiral segregation in the nematic phase, but has two chiral and two achiral configuration of about equal energy~\cite{Vij2023}. All in all, we do not expect explicitly allowing for non-chiral conformers to alter the physics considerably. We intend to investigate this in future work.

\begin{acknowledgments}
This work used the Delta supercomputer at the National Center for Supercomputing Applications through allocation PHY240256 from the Advanced Cyberinfrastructure Coordination Ecosystem: Services \& Support (ACCESS) program, which is supported by U.S. National Science Foundation grants \#2138259, \#2138286, \#2138307, \#2137603, and \#2138296. This research was supported by Grant No. 2022197 from the United States-Israel Binational Science Foundation (BSF). We  thank Dr. Mark Wilson (Durham University, UK) and Dr. Daniela Cywiak (University of Guanajuato, Mexico) for stimulating discussions.
\end{acknowledgments}

\bibliography{references}

\newpage\hbox{}\thispagestyle{empty}\newpage
\onecolumngrid
\appendix
\renewcommand{\thefigure}{S\arabic{figure}}
\setcounter{figure}{0} 

\section*{Supplementary Information}

\subsection*{{Dopant interaction with mesogen enantiomers}}
A simple model for the {interaction} of chiral dopants to achiral mesogens may be provided by a Flory-Huggins type of fluid lattice model. The total number of lattice sites equals $M$. Let $\phi$ be the volume fraction of free mesogens, and $\phi_*$  and $\phi_2$ that of free dopant molecules in the fluid and {dimeric complexes} of mesogens and dopants, respectively. {In reality, these dimers are not necessarily chemically (covalently) bound dimers but may in fact involve interaction complexes of a single dopant to more than a single nematogen. If so, then this merely renormalizes the binding free energy between dopant and mesogen to be introduced below. The free dopants interact much less strongly with the mesogens than those involved in what we refer to as the complexes.} The volume fractions of all species add up to unity, $\phi+\phi_*+\phi_2=1$. Our model dimers take up two lattice sites, whilst free mesogens and free dopants each occupy only a single lattice site. The overall volume fraction of free and bound dopants is $\Phi_*$ and that of the free and bound mesogens is $\Phi = 1 - \Phi_*$. Hence, we have $\phi_*=\Phi_*-\frac{1}{2}\phi_2$ and $\phi=\Phi-\frac{1}{2}\phi_2$.

{For the purpose of evaluating the net interaction free energy between strongly interacting mesogens and dopants, we may ignore any net attractive interaction between mesogens, dopants and dimers, and account only for their excluded volumes implying that the mesogen fluid acts as an ideal solvent to the dopants and the dimers. We do account for a binding free energy $-g/k_\mathrm{B}T<0$ associated with each interaction complex.} The Helmholtz free energy $F$ of the fluid mixtures in that case reads
\begin{equation}\label{eq:freeenergy}\tag{S1}
    \frac{F}{Mk_\mathrm{B}T}= \phi \ln \phi + \phi_* \ln \phi_* +\frac{1}{2}\phi_2 \ln \frac{1}{2}\phi_2 -\frac{1}{2}\phi_2 \left( \frac{g}{k_\mathrm{B}T }\right),
\end{equation}
where $k_\mathrm{B}T$ is the usual thermal energy, the first three terms account for the ideal entropy of mixing and the last for the binding of mesogens and dopants to form dimers. {Parenthetically, we not that if the dimers involve not one but $n$ mesogens, then the $1/2$ in the third term becomes $1/(n+1)$ and $g$ transforms to $n g$. This is renormalization we referred to above. The thermodynamics complexation does not change appreciably. Practically, it means $g$ is implied to have absorbed in it the unknown number $n$.}

The optimal fraction dimers $\phi_2$ minimizes the free energy Eq.~\ref{eq:freeenergy} ,
\begin{equation}\label{eq:massaction}\tag{S2}
    \frac{\phi_2}{\phi \hspace{0.1cm}\phi_*}= 2\exp\left(\frac{g}{k_\mathrm{B}T}\right) \equiv 2 K,
\end{equation}
where we defined the binding constant $K\equiv \exp (g/k_\mathrm{B}T)$.

It is useful to define the enantiomeric excess $\eta$, where we note that due to the presumed interconversion of P and M or $+$ and $-$ conformers, half the free mesogens are $+$ and half $-$. Let a chiral dopant  bound to a mesogen turn this mesogen into a $+$ conformer. In that case $\phi_+ = \frac{1}{2}\phi + \phi_2$ and $\phi_-=\frac{1}{2}\phi$, and $\eta=(\phi_+-\phi_-)/\Phi = \phi_2/2\Phi$. Making use of the conservation of mass, we obtain from Eq.~\ref{eq:massaction}
\begin{equation}\label{etaequation}\tag{S3}
    \frac{\eta}{(1-\eta)(\Phi_*-\eta \phi)} = K.
\end{equation}
This equation can be solved exactly,
\begin{equation}\label{eta}\tag{S4}
    \eta = \frac{1}{2} (1+r+\varepsilon) - \frac{1}{2}\sqrt{(1+r+\varepsilon)^2-4r},
\end{equation}
where $r=\Phi_*/\Phi$  and $\varepsilon=1/K\Phi$. Notice that in the limit where the dopant concentration goes to zero, and $r = 0$, we insist that the enantiomeric excess goes to zero too, $\eta = 0$, in agreement with Eq.~\ref{eta}. In the limit of small dopant concentrations $r\ll 1$, we may Taylor expand Eq.~\ref{eta} to give
\begin{equation}\label{eq:etasmallr}\tag{S5}
    \eta \sim \frac{r}{1+\varepsilon}.
\end{equation}

The free energy gain of complexation is equal to $- \frac{1}{2} g \phi_2 = - g \eta \Phi$ per lattice site. This means that the free energy gain per mesogen must be equal to $- g\hspace{0.1cm } \eta$. In the limit $r\ll 1$, we can replace $\eta$  by Eq.~\ref{eq:etasmallr}. In the same limit $r=\Phi_*/\Phi = \Phi_*/(1-\Phi_*) \sim \Phi*$ and $\varepsilon =  1/K\Phi \sim 1/K$. If the binding is strong and $-g \gg k_\mathrm{B}T$, we have $\varepsilon \ll 1$, and the binding free energy per mesogen becomes equal to $-g\Phi_*$ which can still be much smaller than the thermal energy if $\Phi_*$ is sufficiently small. This is the expression that we use in the main text, where we realize that for the lattice model, the volume fraction dopant is equal to its mole fraction, $X = \Phi_*$.  

{In our model, we need not specify what physics underlies the binding free energy $g$, but it is nevertheless useful to spend a few words on it. In reality, the interaction process involves the alignment of one or more mesogens at the surface of the chiral dopant, such that it maximizes the van der Waals interaction with it. This arguably causes a loss of configurational entropy as well as van der Waals interaction with other mesogens as the ``bound'' mesogens may no longer be aligned with the bulk director field that in the model responds to the interaction with the dopant, as discussed in the main text. Part of the configurational entropy loss of the mesogens is that of the breaking of chiral symmetry. This entropy loss is dealt with explicitly in the full theory described in the main text. The dopant itself might also deform or have its conformational fluctuations affected by the interaction with the nematogens. This implies that $g$ has contributions from both the nematogens and the dopants.}

\subsection*{Temperature dependence of the helical twisting power}
According to our Maier-Saupe theory, the dimensionless helical twisting power can be written as
\begin{equation}\tag{S6}\label{eq:beta}
    \beta = \frac{1}{1-\chi}  \left(\frac{q_0 g}{\pi k_\mathrm{B}T}\right), 
\end{equation}
where each term in the product should be expected to depend on the temperature $T$, since $\chi=\chi(T)$, $q_0=q_0(T)$ and $g=g(T)$. For instance, if we expand the last term around a reference temperature $T_0$ to linear order in $\Delta T /T_0= (T-T_0)/T_0$, we get
\begin{equation}\tag{S8}\label{eq:g}
    \frac{g(T)}{k_\mathrm{B}T}=\frac{g(T_0)}{k_\mathrm{B}T_0}-\left(\frac{h(T_0)}{k_\mathrm{B}T_0}\right) \frac{\Delta T}{T_0}+ \cdots,
\end{equation}
where the free energy gain $g=h-Ts$ can be written in terms of the enthalpy gain $h$ and the entropy loss $s$ of binding. For binding to occur we must heave $g>0$. If $h>0$ and $s>0$ then the binding is enthalpy-driven, if $h<0$ and $s<0$ the binding is entropy-driven. If all other factors do not depend on temperature, then $\beta$ decreases with increasing temperature if the binding is entropy driven.

If we now consider the second term, and Taylor expand $q_0$ near $T=T_0$, we obtain
\begin{equation}\tag{S8}
    q_0(T)=q_0(T_0)+\frac{1}{J(T_0)}\left(-\frac{1}{2}S_K(T_0) + q_0(T_0) S_J(T_0))\right) \frac{\Delta T}{T_0},
\end{equation}
where $K=H_K-TS_K$ and $J=H_J-TS_J$ in terms of the enthalpy gains $H_{J,K}(T)$ and entropy losses $S_{J,K}(T)$ associated with the chiral and nematic interaction strengths $K\geq 0$ and $J\geq 0$. If the interaction strengths $J$ and $K$ are purely enthalpic in nature and $S_{J,K}=0$, as is usually presumed, then $q_0$ does not depend on the temperature. 

This is not what is usually seen experimentally albeit that the temperature dependence can be quite weak over a large temperature range, and for some chiral nematogens $q_0$ may in fact switch sign at some temperature~\cite{Dierking2014}. To account for any temperature dependence, the Maier-Saupe theory for cholesterics has to be extended {to include a higher order Legrendre polynomial in the interaction between the nematogens}~\cite{vandermeer1976}. 
We argue, however, that since $J$ and $K$ represent interactions in a coarse-grained model in which microscopic degrees of freedom have been tacitly ``integrated out'', these quantities should be viewed as free energies rather than energies (or enthalpies). Hence, depending on the sign and magnitude of $S_J$ and $S_K$, $q_0$ may increase or decrease with increasing temperature. This is similar in nature to how the Flory-Huggins interaction parameter of polymer solutions acquires its temperature dependence, and may increase or decrease with increase temperature depending on solvent and polymer type, leading to upper and lower critical solution temperatures.

The temperature dependence of the first term is clearly non-trivial, at least in principle, because $\chi=q_0 K S^2/k_\mathrm{B}T = 2 q_0^2 S^2 J/k_\mathrm{B}T$ is a function of a number of temperature dependent quantities. Even if $q_0$ does not appreciably depend on the temperature, the scalar nematic order parameter does, noting that $S=S(J/k_\mathrm{B}T)$ decreases with increasing temperature. Hence, we would expect $\chi$ to decrease with increasing temperature, and with that $(1-\chi)^{-1}$ too. 

In conclusion, we expect $\beta$ to decrease with temperature, unless the binding of the dopant to the mesogens is entropy dominated and/or the magnitude of the pitch $q_0$ of the (hypothetical) enantiomerically pure compound increases with temperature, in which case $\beta$ may increase with temperature. In fact, we cannot exclude the possibility that depending on the specific temperature dependencies of $q_0$, $g$ and $\chi$, $\beta$ it may even vary non-monotonically with temperature as is sometimes seen experimentally~\cite{Dierking2014}.

\subsection*{Supplementary Figures}

\begin{figure}[ht!]
    \centering
    \includegraphics[width=0.9\textwidth]{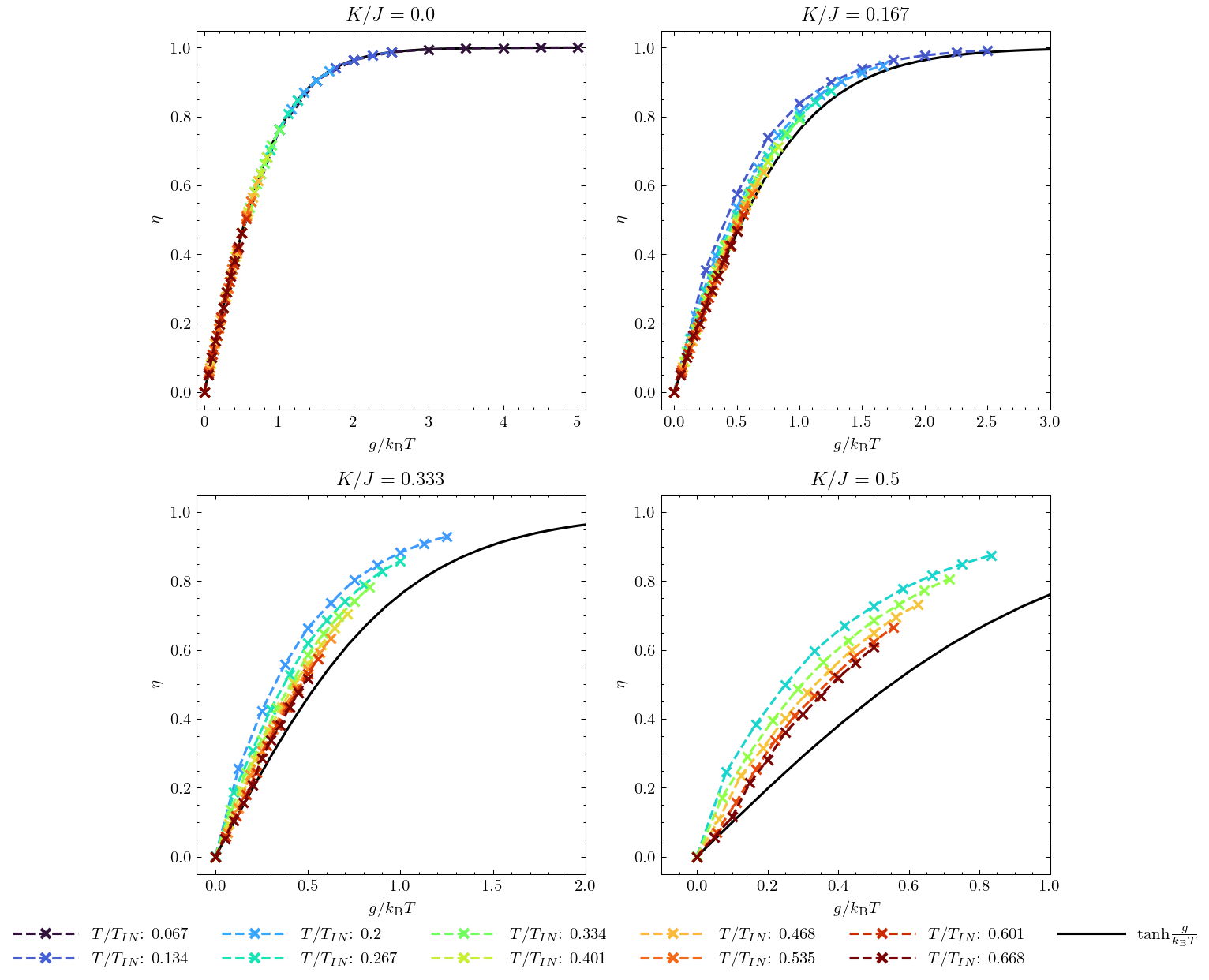}
    \caption{Monte Carlo simulation results showing the enantiomeric excess, $\eta$, as a function of the dimensionless chiral biasing potential $g/k_\mathrm{B}T$. Configurations of $128^3$ sites were annealed from $T/T_{IN} = 0.668$ to $T/T_{IN} = 0.267$ for $2 \times 10^6$ Monte Carlo steps at each temperature. Shown in solid black is the analytical function, $\eta = \tanh \frac{g}{k_\mathrm{B} T}$, representing the limit of zero chiral coupling, $K/J=0$.}
    \label{fig:enantiomeric_excess_multi}
\end{figure}

\begin{figure}[ht!]
    \centering
    \includegraphics[width=0.9\textwidth]{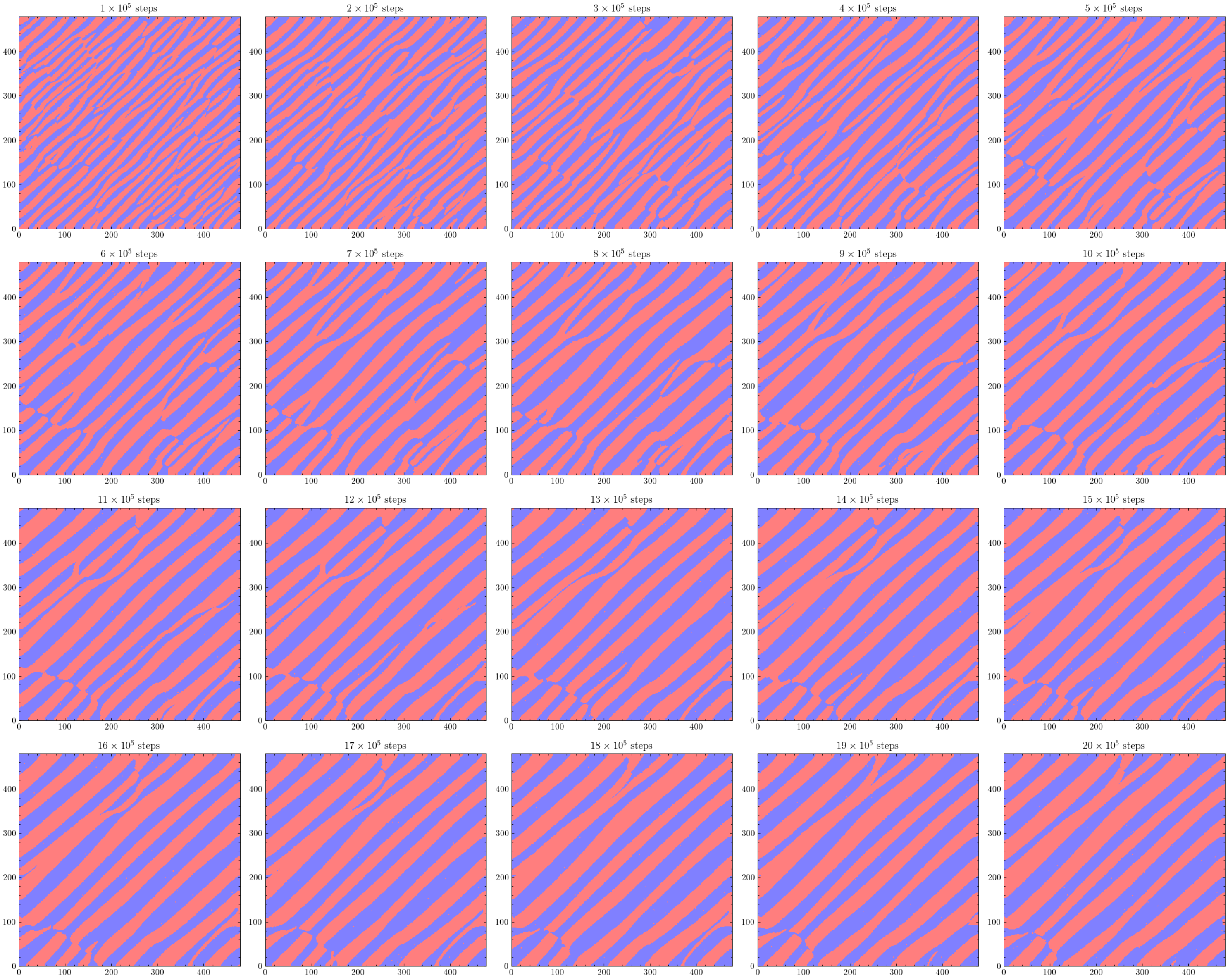}
    \caption{Monte Carlo simulation results showing the coarsening behavior of chiral domains at a constant temperature and chiral interaction strength. Simulations of a box of $480^3$  sites were started in a racemic isotropic state, and run at a constant temperature of $T/T_{IN} = 0.067$ and chiral interaction strength of $K/J = 0.5$. Configurations were saved for visualization purposes every $1 \times 10^5$ Monte Carlo steps.}
    \label{fig:hero_chirality}
\end{figure}
\begin{figure}[ht!]
    \centering
    \includegraphics[width=0.65\textwidth]{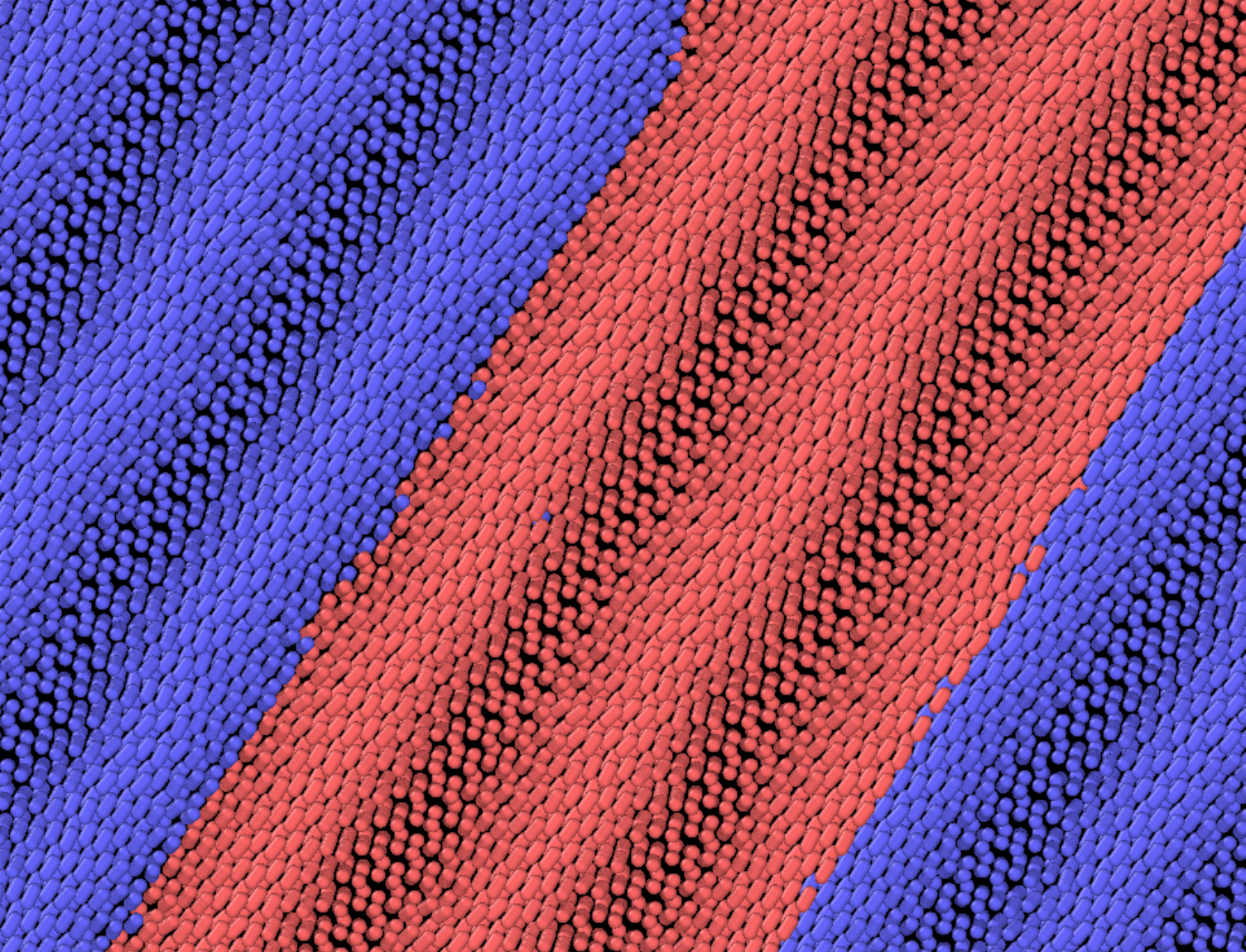}
    \caption{Cropped  cross-section along simulation box diagonal $(+1, -1, +1)$ of a Monte Carlo simulation of size $N=480$ with $K/J = 0.5$ and $T/T_{IN} = 0.067$ after $2 \times 10^6$ Monte Carlo steps showing the director configuration in the deracemized cholesteric phase. The colors red and  blue indicate the left-  and right-handed enantiomeric state of the model mesogens.}
    \label{fig:hero_inset}
\end{figure}


\begin{figure}
\centering
  \begin{subfigure}[b]{0.49\textwidth}
    \includegraphics[width=\textwidth]{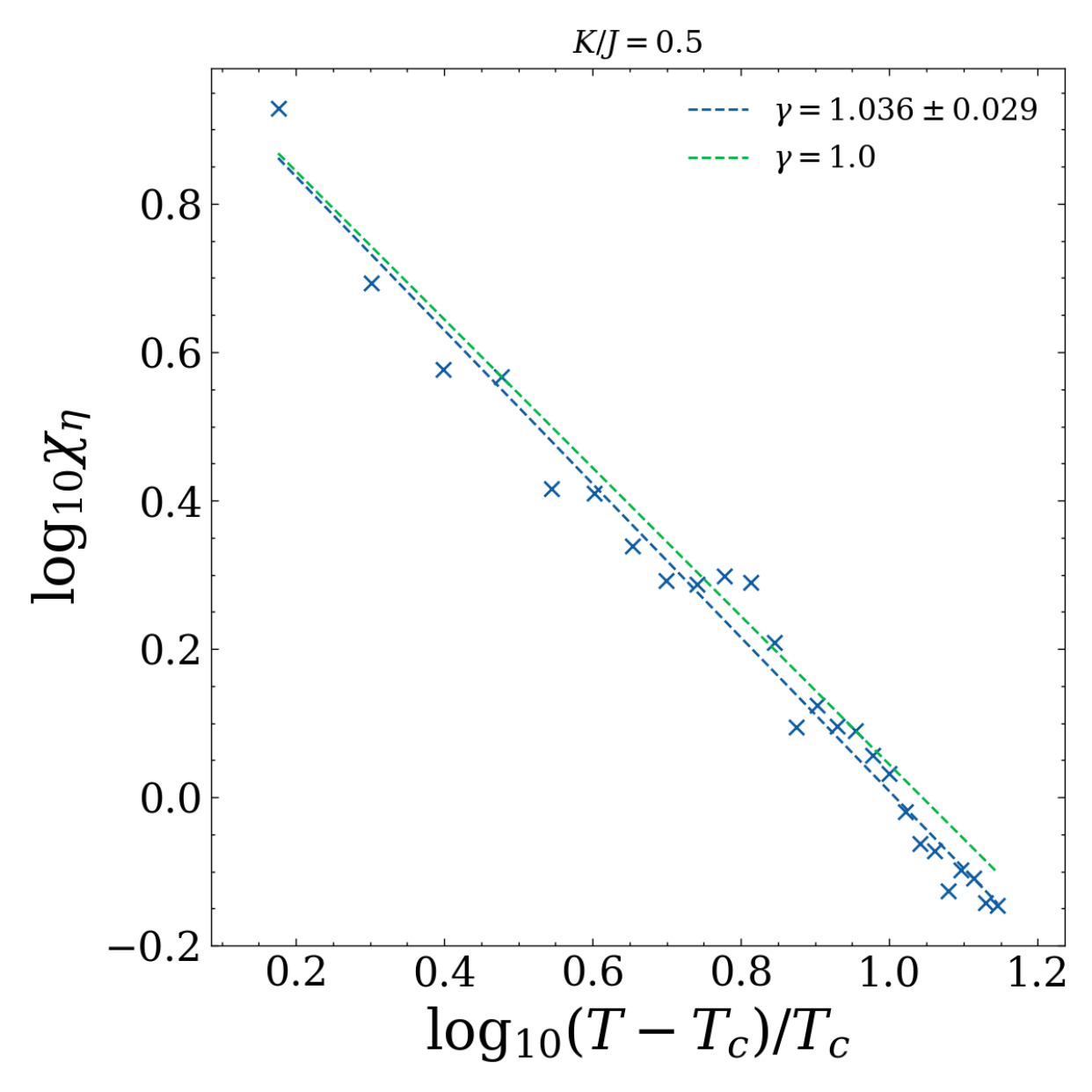}
  \end{subfigure}
 \hfill
  \begin{subfigure}[b]{0.49\textwidth}
    \includegraphics[width=\textwidth]{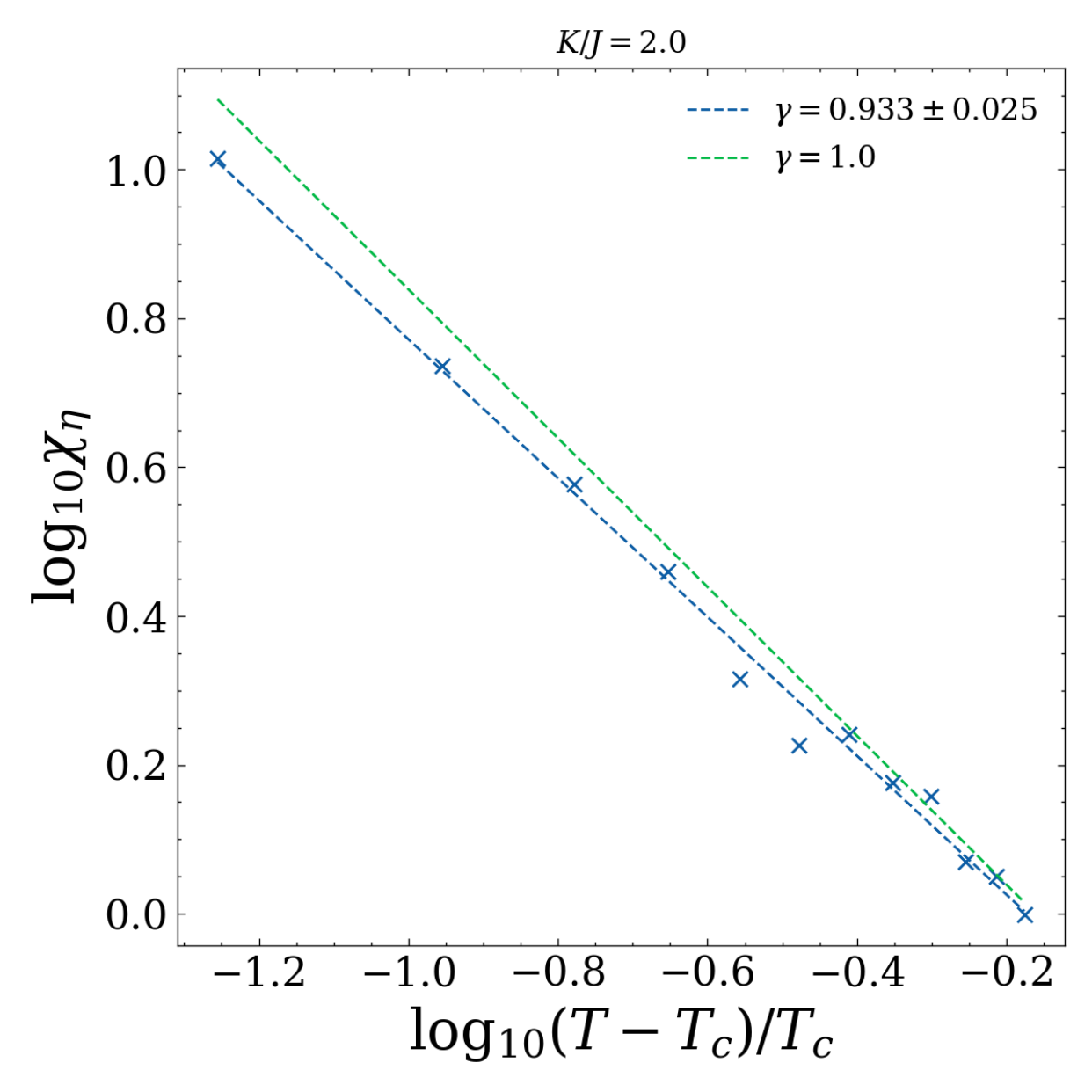}
  \end{subfigure}
    \caption{(Left) Chiral susceptibility in the racemic nematic phase, $\chi_\eta$, for $K/J = 0.5$ and critical transition temperature, $T_\mathrm{c}/T_{IN} = 0.134$. Drawn lines show the mean-field exponent of $\gamma = 1$ and fitted exponent of $\gamma = 1.036$. (Right) Chiral susceptibility in the racemic nematic phase, $\chi_\eta$, for $K/J = 2.0$ and critical transition temperature, $T_\mathrm{c}/T_{IN} = 1.2$. Drawn lines show the mean-field exponent of $\gamma = 1$ and fitted exponent of $\gamma = 0.933$. }%
    \label{fig:chiral_sus_tc}%
\end{figure}

\begin{figure}[ht!]
    \centering
    \includegraphics[width=\textwidth]{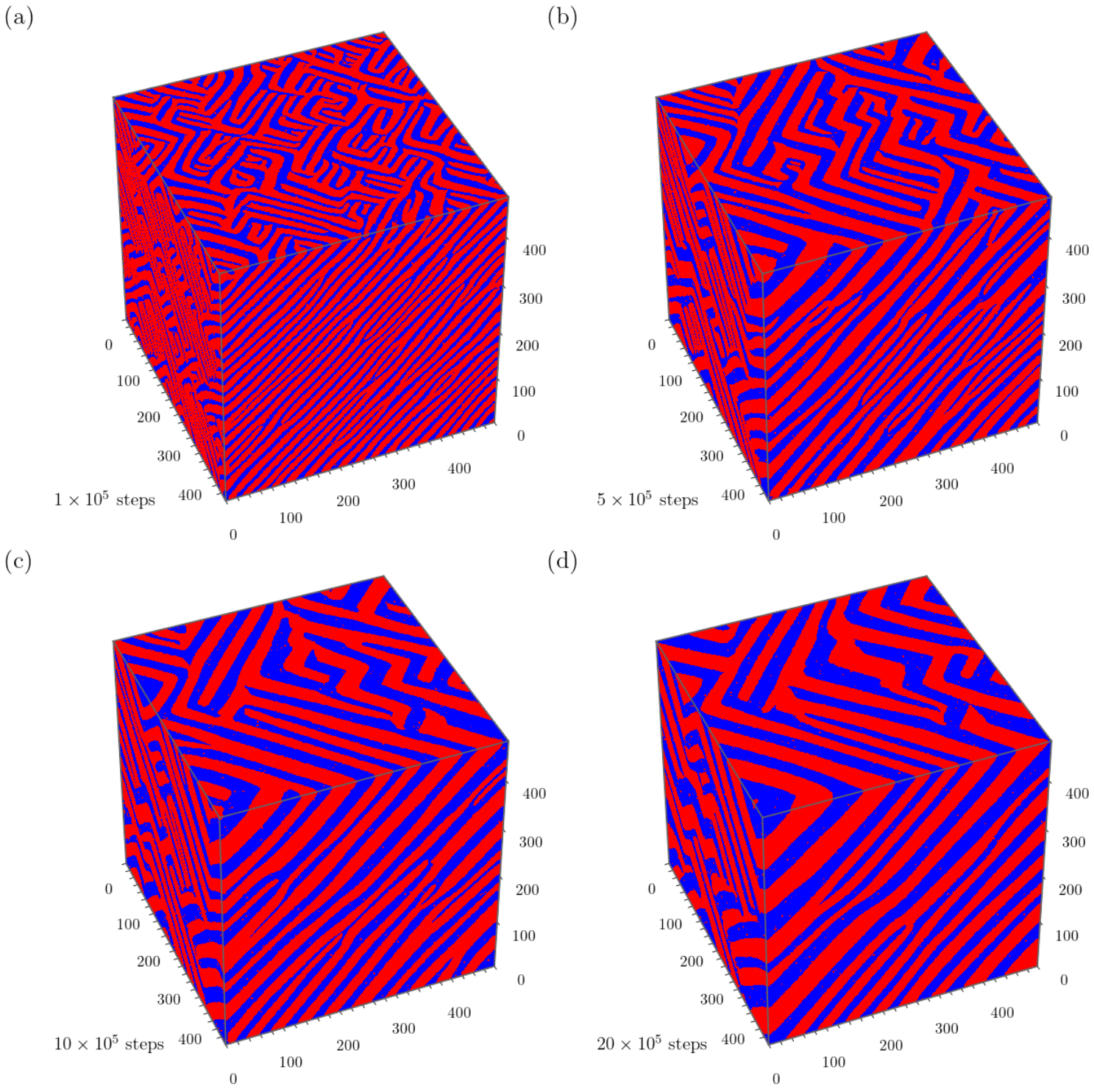}
    \caption{Monte Carlo simulation results showing the coarsening behavior of chiral domains at a constant temperature and chiral interaction strength. Simulations of size $480^3$ were started in the racemic isotropic state and run at a constant temperature of $T/T_{IN} = 0.067$ and chiral interaction strength of $K/J = 0.5$ with configurations saved for visualization every $1 \times 10^5$ Monte Carlo steps.}
    \label{fig:3d_hero}
\end{figure}

\begin{figure}[ht!]
    \centering
    \includegraphics[width=\textwidth]{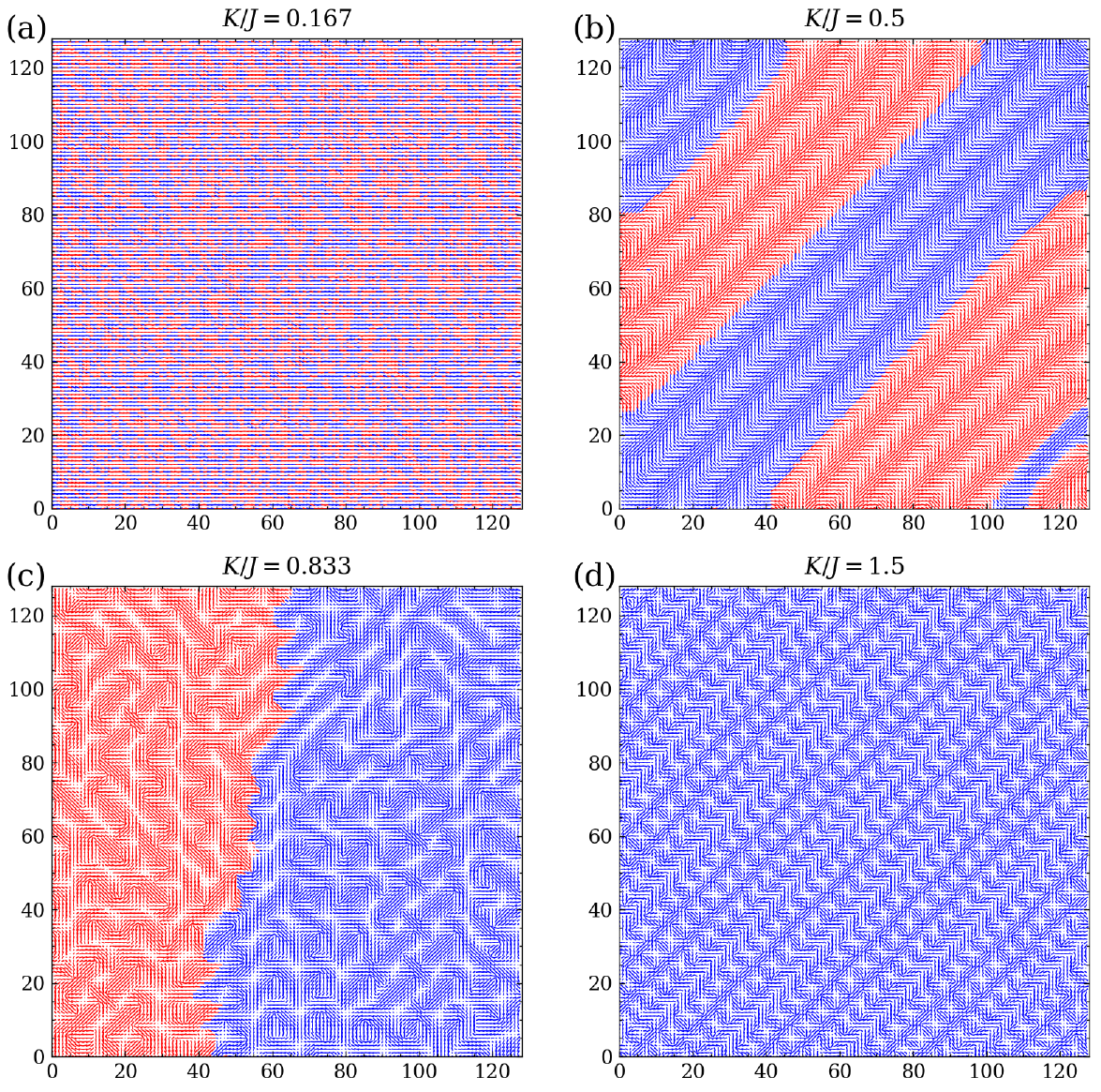}
    \caption{Monte Carlo simulation results showing the final director, and chirality configurations at various values of the strength of the chiral interaction between nematogens. All simulations were annealed to a temperature of $T/T_{IN} = 0.067$ from $T/T_{IN} = 2.0$ in $30$ steps for $2 \times 10^6$ Monte Carlo steps for a total of $6 \times 10^7$ steps.}
    \label{fig:chiral_stripes}
\end{figure}

\begin{figure}
    \centering
    \includegraphics[width=0.4\textwidth]{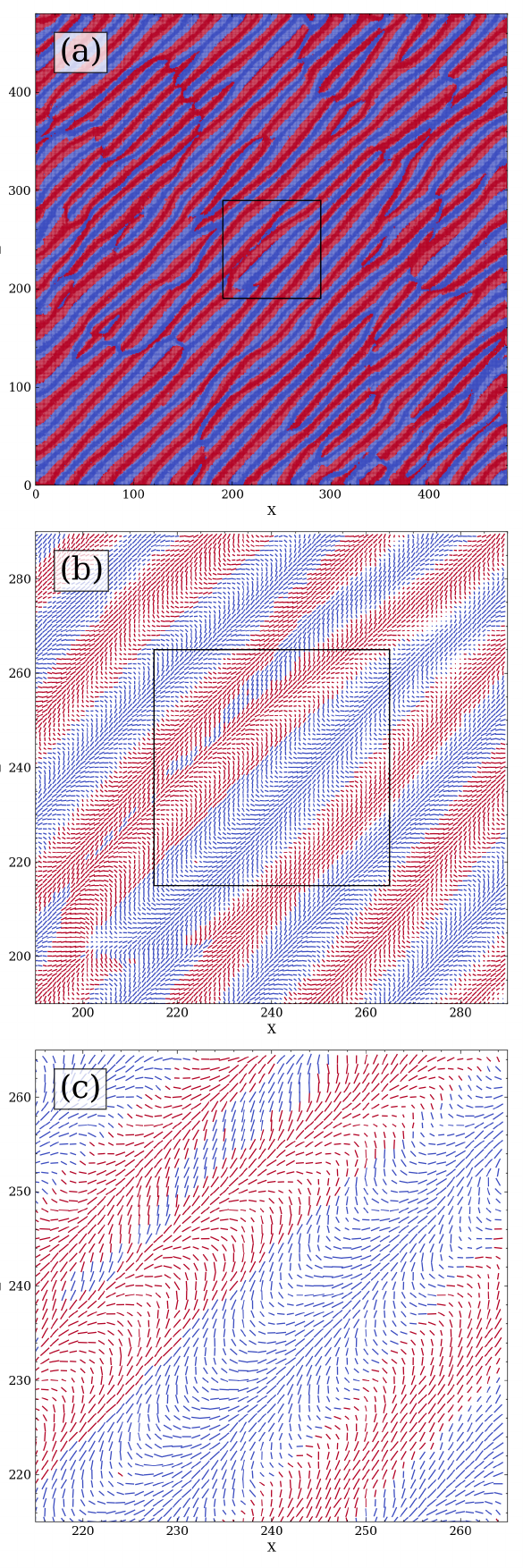}
    \caption{{(a) Monte Carlo simulation results showing the director and chirality configuration of a $N = 480$ simulation along the $XZ$ axes at the midpoint of the $Y$-axis, with $K/J = 0.5$ and $T/T_{IN} = 0.067$ after $1 \times 10^5 $ Monte Carlo time steps. (b) Cropped inset showing a $100 \times 100$ region of the $N = 480$ simulation shown in (a) by the outlined region. (c) Cropped inset showing a $50 \times 50$ region of the $N = 480$ simulation shown in (b) by the outlined region.}}
    \label{fig:S6}
\end{figure}

In this section we present additional material obtained from our simulations.

Fig.~S1 presents additional plots of enantiomeric excess as a function of $g/k_\mathrm{B}T$. We include simulations with zero chiral coupling, $K/J = 0.0$, to shown that as chiral interactions decrease, the enantiomeric excess approaches  the analytical function, $\eta = \tanh \frac{g}{k_\mathrm{B} T}$. The enantiomeric excess was calculated from the last $6.25\% $ of each simulation's annealing step.

Additional figures supporting the discussion of coarsening are shown in Fig.~S2. We show the same, large, simulation of size $N=480$, but visualize the chiral domains through a cross-section, showing the $XZ$ plane through the middle of the simulation volume, with the colors blue and red representing the value of the Ising-like chirality variable describing left- and right-handedness of the model mesogens. We can see that, over time, chiral layers widen and the number of layers drop, that the layer normal is along the diagonal, and that edge dislocations between layers are recombining towards the open boundaries of the simulation volume.

Fig.~S3 shows an inset of the $N=480$ simulation in its final configuration where the chiral layers are the widest. We visualize both the nematic director as a sphero-cylinder and the chiral state variable in color. The helical arrangement of the director field of the cholesteric is clearly shown, as is the sharp interface between the left- and right-handed cholesteric domains. We can also see that there are multiple pitches in each chiral domain, as well as a double-twist structure.

Fig.~S4 shows {two double-logarithmic plots of the chiral susceptibility, $\chi_\eta \sim |T-T_\mathrm{c}|^{-\gamma}$ as a function of the temperature $T -T_\mathrm{c}$ relative to the critical temperature. For the left plot, $K/J = 0.5$, yielding critical temperature $T_\mathrm{c}/T_{IN} = 0.134$ which is located in the racemic nematic phase. The mean-field exponent of unity from fit of simulation data is shown in green and the fitted exponent is shown in blue. The best fit gives a value of $\gamma=1.036\pm 0.029$. The right plot shows a similar fit above the critical endpoint, at $k/J = 2.0$ which has a critical temperature $T_\mathrm{c}/T_{IN} = 1.2$. The critical exponent $\gamma = 0.933\pm0.025$ is less than unity, suggesting that we cannot make a determination on whether the phase diagram contains a critical endpoint or a tricritical point, as $\gamma$ differs above and below this transition.}

Here, in Fig.~S5, we continue to show the coarsening behavior of the large, $N=480$ simulation, visualized on the outer layer of chiral domains as shown in Fig.~4 c). The most prominent outer face shows that the chiral layer normal is along a body diagonal, while the energetically favorable axis is more complex along other faces such as the top. Layers may annihilate at the corners, due to the open boundary conditions allowing them to reduce their surface area easily here. This effect can especially be seen between Fig.~S5 b-d).

Fig.~S6 shows final director and chirality configurations for completely annealed $N=128$ systems in various locations of the phase diagram. Fig.~S6 a) is in the racemic nematic phase and, while it shows small clusters of like chirality, there is no long-range chiral ordering and the nematic director is clearly visible. Fig.~S6 b) shows a system that has just transitioned into a deracemized cholesteric phase with the chiral domains encompassing multiple pitches. Fig.~S6 c) shows a system that hints at transitioning to the deracemized cholesteric blue phase as shown in Fig.~S6 d). All director and chirality configurations were visualized as the horizontal cross-section through the middle of the simulation volume on the $XZ$ plane with director field colored with the value of the Ising-like chiral variable.

{Fig.~S7 shows the director and chirality configurations for a $N=480$ simulation. The interface between left- and right-handed cholesteric layers is sharp, even after only $1\times10^5$ time steps, and is shown by insets focused on the layers far from the open boundaries. By examining the conditions away from the boundary, we can see that the open boundary conditions do not have a significant effect on the layer formation, nor the selection of the cholesteric pitch axis.}

\subsection*{Supplementary Video}
Video~S1 shows an animation of the outer layer of chiral domains, as shown in Fig.~S5 and Fig.~4c), to better show the coarsening behavior and the annihilation of layers of alternating chirality. This animation is of a singular simulation of size $480^3$ started in the racemic isotropic state and run at a constant temperature of $T/T_{IN}=0.067$ without annealing and a constant chiral interaction strength of $K/J=0.5$. Configurations of both the director and the Ising-like chiral variable were saved every $1\times10^5$ Monte Carlo steps per spin for a total of $2\times10^6$ steps and 20 frames of visualization.

\FloatBarrier
\subsection*{Chiral susceptibility of 5CB}
The nematogen 5CB has an isotropic-nematic transition temperature of $T_\mathrm{IN}=307$ K, and a nematic-to-crystal transition at $T_\mathrm{Cr}=296$ K~\cite{Goodby2024}. This implies that the nematic is stable over a temperature range of $11$ K. Hence, for the chiral transition to be observed, $T_\mathrm{c}/T_\mathrm{IN}$ must be larger than $0.96$. Since it has not been observed, $T_\mathrm{c}/T_\mathrm{IN}$ must be smaller than this value. In Figure S8 we have plotted values of the chiral susceptibility $\chi_\eta$ as a function of the scaled temperature $T/T_\mathrm{IN}$ for a number of imagined values of $T_\mathrm{c}/T_\mathrm{IN}$ around the value of unity. Indicated with the dashed line is the crystal transition temperature. All curves diverge upon approach of the chiral transition temperature as $|T-T_\mathrm{c}|^{-1}$. Note that even if $T_\mathrm{c}<T_\mathrm{IN}$, the chiral transition still contributes to the chiral susceptibility and hence to the helical twisting power. For a discussion, we refer to the main text.
\begin{figure}[ht!]
    \centering
    \includegraphics[width=0.65\textwidth]{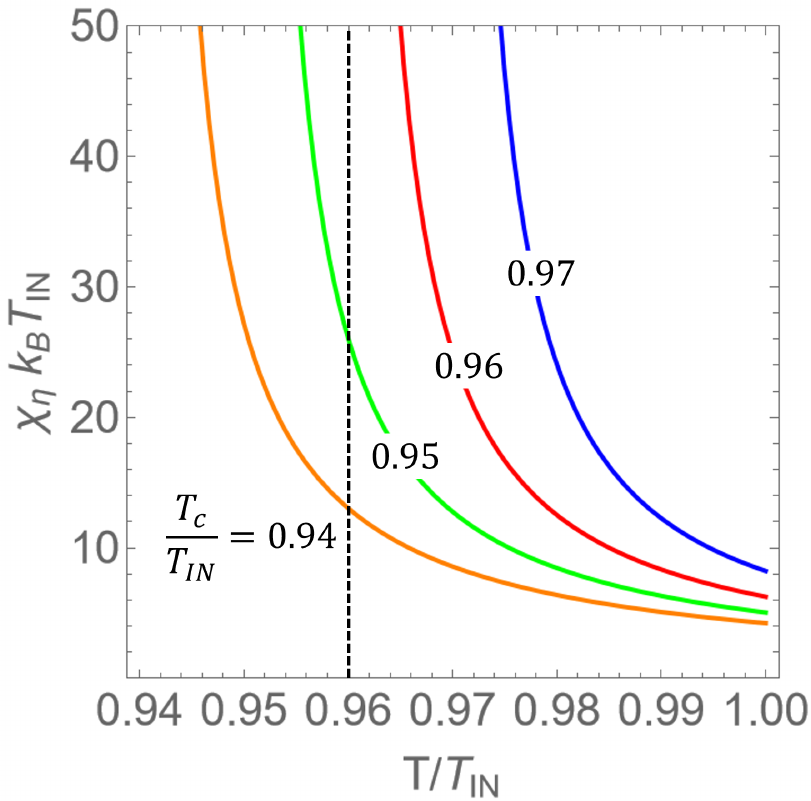}
    \caption{{Dimensionless chiral susceptibility $\chi_\eta k_\mathrm{B}T_\mathrm{IN}$ as a function of the scaled temperature $T/T_\mathrm{IN}$, for different indicated values of the scaled chiral transition temperature $T_\mathrm{c}/T_\mathrm{IN}$. The dashed line indicates the crystal transition temperature of the compound 5CB.  $T_\mathrm{IN}$ is the isotropic-nematic transition temperature, $T_\mathrm{c}$ the chiral transition temperature.} }
    \label{fig:S8}
\end{figure}

{
\section*{Helical twisting power: predictions compared}
The prediction for the dimensionless helical twisting power $\beta$ from our Maier-Saupe theory is given in equation~\ref{eq:beta}. Here, we seek comparison with the theory of Ferrarini and collaborators~\cite{Ferrarini1995,Ferrarini1996,Ferrarini1996(2)}, which has been used to predict the helical twisting power of a class of chiral dopants from computer simulations~\cite{Earl2003}. The expression can be cast into the following form:
\begin{equation*}\tag{S9}\label{eq:betaferrarini}
  \beta = \left(\frac{W \mathcal{C}}{2\pi K_{22}}\right),  
\end{equation*}
Here, $W$ can be seen interpreted as an anchoring surface energy of the director field of the nematic to the surface of the dopant, $\mathcal{C}$ a chirality order parameter describing the the coupling between the chiral surface of the molecule and its orientational ordering in the nematic host, and $K_{22}$ the twist elastic constant of the nematic host. The expression has been made non-dimensional by some unimportant microscopic length scale.}

{Arguably, $W$ and $\mathcal{C}$ depend on both the dopant and the nematogen, while $K_{22}$ is a property of the nematic host fluid. This suggests the following correspondence between both models: $W\mathcal{C}\leftrightarrow g$. The correspondence between $(1-\chi)^{-1} q_0$ and $1/K_{22}$ of Eqs.~\ref{eq:beta} and \ref{eq:betaferrarini} is less obvious, as in the model of Ferrarini and collaborators the nematic is treated as a deformable background. However, since $q_0=K/2J$, and since $K_{22} \propto J$~\cite{Revignas2024}, we do have $q_0\propto 1/K_{22}$. This is reassuring, and shows that both approaches have at least some common ground.}

{Clearly, both the factor $\chi \propto K^2$ as well as the $q_0 \propto K$ involve the chiral interaction strength $K$ that is absent in the usual description of nematics. It is important to realize that the chiral shape theory is inherently macroscopic in nature, whilst our chiral fluctuation theory is microscopic in nature and does explicitly describe interaction of a director field with a chiral surface. It means that in the limit where $K\rightarrow 0$, our theory produces a zero helical twisting power. This, obviously, is drawback of our approach. Ideally, both theories should somehow be merged. } 

{The simplest way to do that seems to be to replace $K_{22}$ in the macroscopic theory by a renormalized one that accounts for the chiral switching of the nematogens. Preliminary calculations show that within Maier-Saupe theory chiral interactions lower the twist constant according to $K_{22}\propto (1-\chi)$, and hence that the twist constant vanishes at the chiral transition temperature. This brings the two theories closer together. We intend to report on this in the near future~\cite{vanderSchoot2025}.}

{\section*{Hysteresis}
The Maier-Saupe model exhibits hysteresis for positive and negative values of the chiral biasing field $g$, provided the chiral interaction parameter $\chi$ is larger than unity, as shown in Figure~\ref{fig:hysteresis}. This is is indicative of a phase transition taking place for $\chi =1$. For $\chi >1$, both $\eta$ and $x$ change sign at zero field strength, $g=0$. The spinodals are state points at values of $g$ for which derivatives of these quantities diverge. The state points between the spinodals and the values at $g=0$ for which the slope is negative are thermodynamically unstable, and those for which the slopes are positive are metastable. The existence of metastable state points gives rise hysteresis phenomena similar to that found in ferromagnets.
\begin{figure}[ht!]
    \centering
    \includegraphics[width=1.0\textwidth]{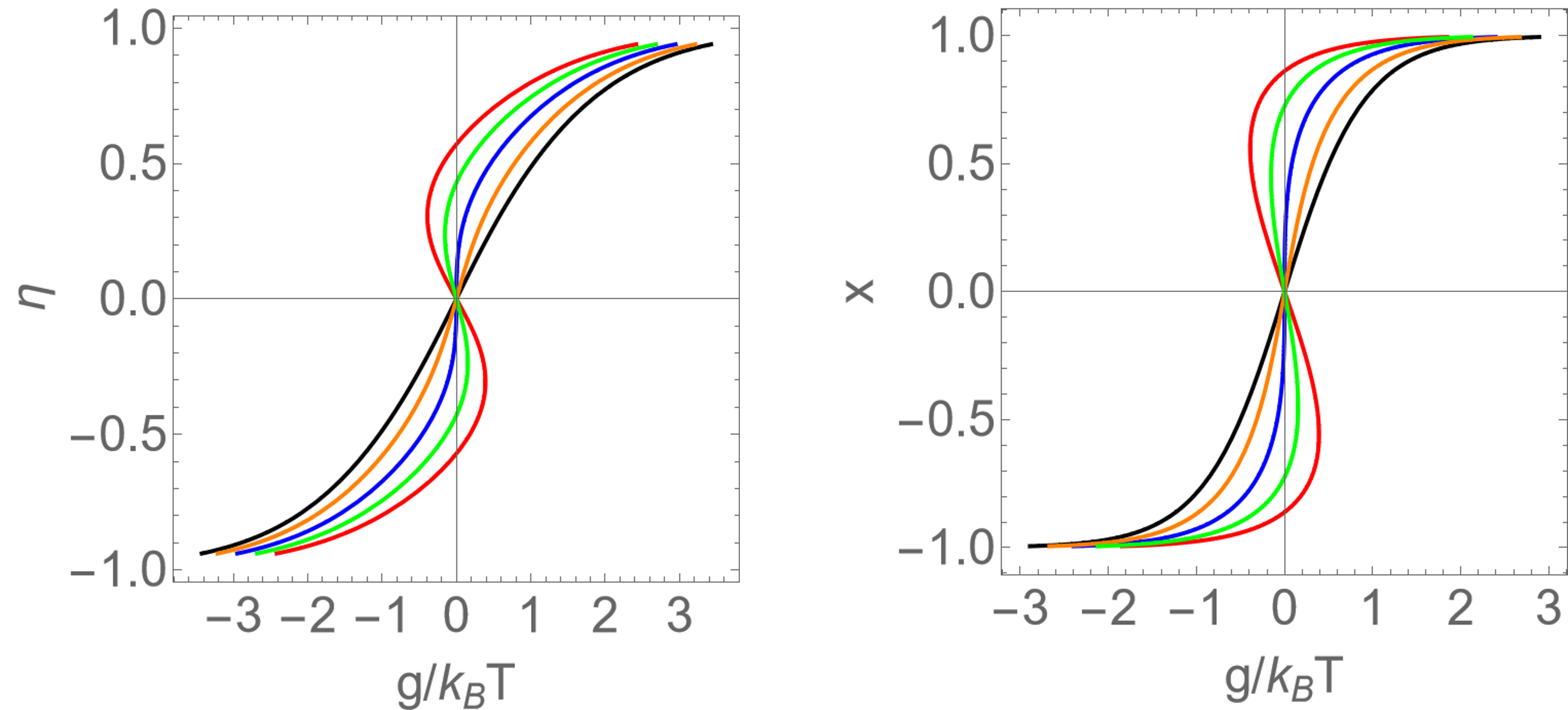}
    \caption{Chiral order parameter $\eta$ (left) and scaled cholesteric wave number $x=q/q_0$ (right) as a function of the dimensionless field strength $g/k_\mathrm{B}T$. Lines indicate state points for which the free energy is stationary. Red: $\chi= 2.0$ ; green: $\chi= 1.5$; blue: $\chi= 1.0$; orange: $\chi= 0.5$; black: $\chi= 0.1$ }
    \label{fig:hysteresis}
\end{figure}
\end{document}